\documentclass[a4paper,11pt]{article}
\pdfoutput=1
\usepackage{jheppub}
\usepackage{amsmath,amssymb}
\usepackage{bbold}
\usepackage{graphicx}
\usepackage{float}
\usepackage [utf8]{inputenc}
\usepackage{subfigure}
\usepackage{euscript}
\usepackage{placeins}

\allowdisplaybreaks[2] 

\def\OO{\mathcal{O}}
\def\bb{{\boldsymbol b}}
\def\bx{{\boldsymbol x}}

\def\be{\begin{equation}}
\def\ee{\end{equation}}
\def\cb{c_\mathrm{B}}
\def\gsqa{g_{3\mathrm{d}}^2a}
\def\gsq{g_{3\mathrm{d}}^2}
\def\gfour{g_{3\mathrm{d}}^4}

\def\ycrit{y_{\mathrm{crit}}}
\def\Tr{\mathrm{Tr}}
\def\Tphilat{\Tr\, \Phi^2_{\mathrm{L}}}
\def\Tphicont{\frac{\Tr\, \Phi^2_{\mathrm{cont}}}{\gsq}}
\def\phisqcont{\frac{\Tr \, \Phi^2}{\gsq}}
\def\phisqcontsymm{\frac{\Tr \, \Phi^2_\mathrm{symm}}{\gsq}}
\def\phisqcontbrok{\frac{\Tr \, \Phi^2_\mathrm{brok}}{\gsq}}
\def\delphisqcont{\frac{\Delta \Tr \, \Phi^2}{\gsq}}

\title{Full $\OO(a)$ improvement in electrostatic QCD}
\author{Guy D.\ Moore, Niels Schlusser}
\affiliation{Institut f\"ur Kernphysik, Technische Universit\"at Darmstadt\\
Schlossgartenstra{\ss}e 2, D-64289 Darmstadt, Germany}
\emailAdd{guy.moore@physik.tu-darmstadt.de, \\ nschlusser@theorie.ikp.physik.tu-darmstadt.de}

\abstract{
  EQCD is a 3D bosonic theory containing SU(3) and an adjoint scalar,
  which efficiently describes the infrared, nonperturbative sector of
  hot QCD and which is highly amenable to lattice study.  We improve
  the matching between lattice and continuum EQCD by determining the
  final unknown coefficient in the $\OO(a)$ matching, an additive
  scalar mass renormalization.  We do this numerically by using the
  symmetry-breaking phase transition line of EQCD as a line of
  constant physics.  This prepares the ground for a precision study of
  the transverse momentum diffusion coefficient $C(q_\perp)$ within
  this theory.  As a byproduct, we provide an 
  updated version of the EQCD phase diagram. }


\begin{document}
\maketitle

\section{Introduction}
\label{sec:intro}

At low energy scales, and therefore at low temperatures, the coupling
of QCD becomes large and the theory's behavior becomes
nonperturbative.  Therefore we should not be surprised if perturbation
theory does not work for thermodynamical or dynamical properties as
one approaches the QCD crossover temperature,
$T \sim 150\:\mathrm{MeV}$ \cite{Bazavov:2014pvz,Borsanyi:2013bia},
from above.  However, it came as a
surprise just how badly perturbation theory works at scales up to many
times the transition temperature.  For instance, thermodynamical
properties such as the pressure have an expansion in $g$ the strong
coupling which is known up to $g^6 \ln(g)$
\cite{Shuryak:1977ut,Kapusta:1979fh,Toimela:1983407,Arnold:1994eb,Zhai:1995ac,Kajantie:2002wa},
and while the leading-order behavior is within $30\%$ of the lattice
result above $360\:\mathrm{MeV}$
\cite{Bazavov:2014pvz,Borsanyi:2013bia},
the series of corrections does not converge even for
$T=100\:\mathrm{GeV}$, a scale where perturbation theory should work
well \cite{Laine:2003ay}.  This problem was first understood broadly
by Linde \cite{Linde:1980ts}, and was diagnosed more completely
starting in the mid-1990s with the work of Braaten and Nieto
\cite{Braaten:1995cm}, who showed
that the perturbative expansion could be understood as a two-step
process.  Treating the problem in Euclidean space, the time direction
is periodic with period $\beta = 1/T$, corresponding in
frequency space to a tower of discrete frequencies
$\omega_n = 2\pi nT$, the Matsubara frequencies (fermions have
$\omega_n = (2n+1)\pi T$).  One can integrate out all but the $n=0$ modes
of the spatial gauge field $A_i$ and its temporal component $A_0$:
dimensional reduction \cite{Appelquist:1981vg,Nadkarni:1982kb}.
This results in a 3-dimensional theory of an SU(3) gauge field and an
adjoint scalar ($A_0$ behaves as a scalar and we will call it $\Phi$
henceforth), which has been christened ``3D Electrostatic QCD'' or EQCD for
short.

Explicit loopwise-order calculations of the matching between full
thermal QCD with quarks and EQCD
\cite{Kajantie:1997tt,Laine:2005ai}
indicate that integrating out the nonzero Matsubara frequencies is a
well-behaved perturbative procedure down to temperatures of order
$300\:\mathrm{MeV}$.  It is the behavior of EQCD itself which is not
under perturbative control.  But one can solve EQCD nonperturbatively
on the lattice, and this appears to generate much closer
approximations to the full 4D thermodynamics than perturbation theory
alone \cite{Laine:2005ai,Hart:2000ha,Hietanen:2008xb}.

EQCD is a 3D theory of bosons only, which is relatively easy to treat
on the lattice; for most thermodynamical quantities, high-quality
results were already available 20 years ago.  But in 2008 Caron-Huot
showed \cite{CaronHuot:2008ni} that EQCD is also the right effective
theory for computing a \textsl{dynamical} transport property,
$C(q_\perp)$ the differential rate at which a highly relativistic
colored particle exchanges transverse momentum with the medium
\cite{Bass:2008rv,CasalderreySolana:2007qw}.
This has important applications in hard particle suppression and jet
modification in the hot QCD medium.  For instance, the frequently
discussed ``transport coefficient'' $\hat{q}$ is defined as the
$q_\perp^2$ moment of $C(q_\perp)$,
$\hat{q} = \int \frac{d^2 q_\perp}{(2\pi)^2} q_\perp^2 \: C(q_\perp)$.
To calculate $C(q_\perp)$, one notes that its transverse-space Fourier
dual $C(\bb_\perp)$ is defined in terms of a Wilson loop with two
spatial and two lightlike edges \cite{CasalderreySolana:2007qw}:
\begin{equation}
  \label{Cdef}
  C(\bb_\perp) = \lim_{L \to \infty} - \frac{1}{L} \ln W[(0,0,0);(0,0,\bb_\perp);
    (L,L,\bb_\perp);(L,L,0)] \,,
\end{equation}
where $W(a;b;c;d)$ is the (fundamental representation) Wilson loop
connecting the four spacetime points $(a;b;c;d)$, and we have written
$a=(a_t,a_z,\boldsymbol a_\perp)$.  What Caron-Huot showed is that $W$
can be replaced with a similar Wilson loop in EQCD, with the
$L$-length edges replaced by a modified Wilson line which also
incorporates the scalar field $\Phi$, up to corrections which should
be highly suppressed in any regime where dimensional reduction works.
It would be extremely interesting to pursue a high-quality
determination of $C(q_\perp)$ within EQCD, to extract
\textsl{directly} important nonperturbative \textsl{dynamical}
information about the behavior of hot QCD, of direct relevance for
experiment.

Panero, Rummukainen, and Sch\"afer have made a first exploration of
this quantity in EQCD \cite{Panero:2013pla}.  However it appears that
the numerics are much more challenging than for standard thermodynamical
quantities, especially for small $\bb_\perp$ values.  Therefore it is
essential to minimize lattice spacing errors, since the numerical cost
of a study increases as approximately $a^{-5}$, with $a$ the lattice
spacing.  In a 3D theory it should be possible to perform a
lattice-continuum matching which is free of $\OO(a)$ errors.
Without doing so, the leading errors are of order
$\OO(a/b_\perp)$; with $a$-errors removed, the leading errors are
$\OO(a^2/b_\perp^2)$, a very significant improvement for small $b_\perp$.
At the level of the Lagrangian parameters of EQCD, such an improvement
was performed a long time ago \cite{Moore:1997np} -- \textsl{except}
that one parameter remains undetermined at the $\OO(a)$ level.
When evaluating a particular composite operator such as the modified
Wilson line, one must also determine the $\OO(a)$ corrections to the
operator; but this was done for the modified Wilson line operator a few years
ago \cite{DOnofrio:2014mld}.  So we would have everything we need to
proceed with a lattice study, free from any $\OO(a)$ errors, if we
only knew the lattice-continuum matching of that one remaining
Lagrangian parameter. Specifically, while most Lagrangian terms
receive $\OO(a)$ corrections at one loop, the $\Phi$-field
mass-squared parameter receives (known) $\OO(1/a)$ corrections at one
loop and $\OO(\ln(a))$, $\OO(1)$ corrections at two loops
\cite{Kajantie:1997tt}.
The unknown $\OO(a)$ errors arise at the 3-loop level
\cite{DOnofrio:2014mld}.

This paper will undertake the necessary technical development of
determining this $\OO(a)$ correction in the lattice-continuum matching
of 3D EQCD theory.
The direct diagrammatic evaluation appears too difficult to pursue;
so we will use an alternative method to extract the corrections.
Since the
$m^2_{3\mathrm{d}}$-renormalization is the only missing
$\OO(a)$-contribution, it can be determined by fitting to a
line of constant physics. EQCD features a phase transition, which we
will utilize to obtain such a line.%
\footnote{%
  Note that the phase transition in EQCD is unphysical in the sense
  that it is not related to any phase transition in 4D thermal QCD.
  Indeed, 4D QCD with physical quark masses has a crossover at zero
  quark chemical potential \cite{Bazavov:2014pvz,Borsanyi:2013bia};
  the pure-glue theory at high temperature has a $Z_3$-breaking phase
  structure which is related to the EQCD phase transition in a 1-loop
  analysis \cite{Kajantie:1997tt}, but because
  EQCD lacks a true $Z_3$ symmetry, this is essentially a coincidence
  (see however Ref.~\cite{Vuorinen:2006nz}).}

In the remainder of the paper, we present our investigation of the
matching problem.  Section \ref{sec:setup} sets the theoretical
stage.  Section \ref{sec:method} presents our approach to determining
the 3-loop mass renormalization indirectly from lines of constant
physics. We present our results in Section \ref{sec:results} and leave
conclusions and outlook to Section \ref{sec:conclusion}.  A few odds
and ends appear in two appendices.

For the impatient reader, here is a very short summary.  The theory
EQCD has two parameters, the mass-squared $y$ and scalar self-coupling
$x$ (the gauge coupling just sets a scale).  For a given value of the
self-coupling $x$, there is a critical $y$-value where a phase
transition occurs.  We find this point at several lattice spacings,
and extrapolate to the continuum behavior; the slope of the fit at
$a=0$ is precisely the $\OO(a)$ mass correction which must be
compensated.  Perturbative arguments show that the resulting slope
should depend on $x$ as a third-order polynomial.
Determining this at several $x$-values allows us to fit all polynomial
coefficients, which are presented with their covariance matrix in
Eq.~(\ref{mainresult}) and Table \ref{cov_mat_grand_fit}.

\section{Theoretical setup}
\label{sec:setup}

The theory EQCD is a 3-dimensional SU(3) gauge theory with gauge field
$A_i^a T^a$ ($i=1,2,3$ and $a=1\ldots 8$) with a real adjoint scalar
field $\Phi$ which can be understood as the dimensional reduction of
the 4D Euclidean $A_0$ field component.  The continuum action is
\begin{equation} \label{cont_action}
S_{\mathrm{EQCD,c}} = \int \mathrm{d}^3x \, \left( \frac{1}{2 \gsq} \Tr \,
 F^{ij} F^{ij} + \Tr \, D^i \Phi D^i \Phi + m_\mathrm{D}^2 \Tr \, \Phi^2 +
  \lambda \big( \Tr \, \Phi^2 \big)^2  \right) \, .
\end{equation}

The parameter $m_D^2$ has logarithmic scale dependence which we
resolve in the same way as in \cite{Kajantie:1997tt}. We will use
the coupling $\gsq$, which has dimensions of energy, to set the scale,
and we work in terms of the dimensionless ratios
$x \equiv \lambda / \gsq$ and
$y \equiv m_D^2(\bar\mu=\gsq)/\gfour$, originally introduced by
\cite{Farakos:1996705,Kajantie:1997tt}.

We will not present the full details of our lattice implementation or
update algorithms, since they are almost identical to
\cite{Kajantie:1995kf}.  We use the standard 
Wilson gauge action and nearest-neighbor scalar gradient or ``hopping'' term.
The only crucial  difference to the presented $SU(2)$ + fundamental scalar-case 
concerns the treatment of the hopping term in the gauge field
update. It arises from the scalar kinetic term, which translates into
\begin{equation}
  \label{latthopping}
  \int \mathrm{d}^3x \, \Tr \, D^i \Phi D^i \Phi \rightarrow 2 Z_\Phi \sum_{\bx, 
i} \Tr \, \left( \Phi_\mathrm{L}^2(\bx) - \Phi_\mathrm{L}(\bx) U_i (\bx) \Phi_
\mathrm{L}(\bx + a \hat{i}) U_i^\dagger (\bx) \right)
\end{equation}
in the lattice formulation, where $\Phi_\mathrm{L}$ is the rescaled, 
dimensionless lattice version of the adjoint scalar field, $Z_\Phi$ is
a field renormalization factor, and $U_i(\bx)$ is the standard gauge link 
at lattice site $\bx$ in direction $i$. In contrast to the fundamental scalar 
case treated in \cite{Kajantie:1995kf}, the present hopping term is non-linear 
in the link. Therefore, it has to be incorporated into the link update via a 
Metropolis step.  Our scalar update, on the other hand, is a mixture
of heatbath updates with the $x \Tr \, \Phi^4$ term included by
Metropolis accept/reject, and the overrelaxation update introduced in
Ref.~\cite{Kajantie:1995kf}.  We update sites in checkerboard order.
Our code was modified from the OpenQCD-1.6 package \cite{openQCD}.

Now we return to the parameters of the continuum and lattice actions.
For this choice of parameters, 1-loop relations between the lattice
gauge and scalar couplings and their continuum values, and two-loop
relations for the scalar mass, are known; we use the expressions from
\cite{DOnofrio:2014mld}.\footnote{%
The paper is written for general gauge groups, where there are two
independent scalar self-couplings.  These are equivalent in $SU(3)$,
so we take $x_2=0$ in their notation. Note that in the lattice action in
\cite{DOnofrio:2014mld}, $x_1$ and $x_2$ actually have to be
interchanged for consistency with the rest of that paper.  Also, since
the normalization of the lattice scalar field is arbitrary, we have
chosen $Z_\Phi=1$, that is, we normalize our hopping term to have unit
norm.}
The matching between the lattice and continuum is such that we know
the lattice $x$ and $\gsq$ parameters up to $\OO(a^2 \gfour)$ corrections.
Effects from higher-dimension operators (present in the Wilson action
and nearest-neighbor hopping term) are also of $\OO(a^2)$.  We
also know the multiplicative rescaling between $y$ and
$y_{\mathrm{latt}}$ to the same precision, and we know the
$\OO(1/a)$ and $\OO(1,\ln(a))$ additive contributions to $y$.  Only
the (3-loop) $\OO(a)$ additive contribution to $y$ is unknown.
Therefore any $\OO(a)$ difference in a physical result between lattice
treatments at different lattice spacings must arise due to this
additive contribution.

The phase structure of EQCD was extensively examined in the 90's, for
example in \cite{Kajantie:1998yc}.  The theory has a $\mathbb{Z}_3$
symmetry which is broken if $\Tr \Phi^3$ takes a nonvanishing value.
There is a line of phase
transitions separating a large-$y$ region, where $\mathbb{Z}_3$
symmetry is preserved, from a small-$y$ region where $\mathbb{Z}_3$
symmetry is spontaneously broken.  Unlike the transition in $SU(2)$
fundamental \cite{Kajantie:1995kf} or adjoint \cite{Kajantie:1997tt} theories, this
transition line extends over all $x$ values, since the phases are
distinguished by a global discrete symmetry breaking.  At small $x$
values the transition is first order; there is a tricritical point,
and for large $x$ values it is second order
\cite{Kajantie:1998yc}.  Values of $x$ corresponding to dimensional
reduction from physical temperatures and quark numbers all land in a
region where the transition is first order; they also lie below the
critical value $\ycrit$, so physical QCD corresponds to metastable
points in the EQCD phase diagram.  (We emphasize again that the phase
transition in EQCD is not related to any thermal phase transitions
which may or may not occur in 4D QCD.)

Our methodology will consist of determining, for a given $x$ value,
the value $\ycrit$ where the phase transition occurs.  Doing so at a
series of lattice spacings provides a lattice determination of the
lattice spacing $a$ dependence of $\ycrit$.  Since the only $\OO(a)$ error
remaining in our lattice implementation of the theory is an additive
shift to $y$, the slope of $\ycrit(a)$ when we extrapolate the lattice
spacing $a\to 0$ determines the unknown linear-in-$a$ correction to
$y$ at each given $x$ value.

Formally, we know that the $\OO(a)$ lattice-continuum additive
$\delta y$ contribution arises from 3-loop scalar self-energy diagrams
in lattice perturbation theory \cite{Moore:1997np}.  Even without
computing these graphs, we can see that they involve 0, 1, 2, and 3
factors of the scalar self-coupling.
Therefore, making $N=3$ in the expression from \cite{DOnofrio:2014mld}
explicit, we give the lattice mass-squared in terms of the continuum
$y$ value, 
\begin{align}
  S_{\mathrm{EQCD,L}} =&  \ldots + \sum_x Z_2 (y+\delta y)
  \Tr\: \Phi^2_{\mathrm{L}} \,, \\
  Z_2 =& \, g_{3\mathrm{d}}^4a^2 \left( 1 + \gsqa \left[ - \frac{\Sigma}{8 \pi} + (- 9 + 10 x) 
  				\frac{\xi}{4 \pi} \right] \right) \\
  \delta y =& -\frac{1}{\gsqa} \frac{\Sigma}{4 \pi} \left( 6 + 10 x \right) 
  				+ \delta y_{3\mathrm{loop}} \notag \\
  			&- \frac{1}{16 \pi^2} \left[ 10 x \bigg( \frac{3}{2} \Sigma^2 + 3 \Sigma \xi - 
  				6 \delta \bigg) + \bigg( \ln \bigg( \frac{6}{\gsqa} \bigg) + \zeta - 3 
  				\Sigma \xi \bigg) \big( 60 x - 20 x^2 \big) \right. \notag \\
  			& \left. \hspace{4em} + 9 \bigg( \frac{7}{8} \Sigma^2 - \frac{\Sigma \pi}{6} 
  				+ \frac{31 \Sigma \xi}{6} + 2 \kappa_1 - \kappa_4 + 4 \rho - 4 \delta 	
  				\bigg) \right] \,,
\end{align}
where $\zeta=0.08849$, $\delta=1.942130$ and $4\rho - 2\kappa_1 + \kappa_4=-1.968325$.
The undetermined $\OO(a)$ additive contribution must be parametrically
of form
\begin{equation}
  \label{parametric}
  \delta y_{3\mathrm{loop}} = \gsqa
  \Big(C_0 + C_1 x + C_2 x^2 + C_3 x^3 \Big) .
\end{equation}
With results at enough $x$ values, we can perform a polynomial fit to
extract these coefficients, and use it to determine the correction at
any $x$ value.

Eventually we want to apply EQCD to study QCD.  Dimensional reduction
at a specific temperature (hence gauge coupling) and number of light
fermions leads to a specific $x$ and $y$ value.  The 2-loop reduction
formulae between high-temperature 3+1 dimensional full QCD and EQCD
were worked out by Kajantie \textsl{et al}
\cite{Kajantie:1997tt,Laine:2005ai} and we use a nonperturbative value of
$\Lambda_{\overline{\mathrm{MS}}}$ from \cite{Bruno:2017gxd}.  These
lead to the specific $x$ and $y$ values, which we will later
investigate for $C(q_\perp)$ behavior, shown in
Table \ref{match_scenarios}.  To minimize errors in a future
investigation, we will examine the mass renormalization at the
$x$-values indicated, except the smallest value where our method will
prove ineffective.  We will also study larger values of $x$ which do
not correspond to any physical QCD regime.

\begin{table}[htbp!] 	
\centering
\begin{tabular}{|c|c|c|c|}	
\hline
$T$ & $n_{\mathrm{f}}$ & $x$ & $y$ \\
\hline
$250 \, \mathrm{MeV}$ & 3 & $0.08896$ & $0.452423$ \\
$500 \, \mathrm{MeV}$ & 3 & $0.0677528$ & $0.586204$ \\
$1 \, \mathrm{GeV}$ & 4 & $0.0463597$ & $0.823449$ \\
$100 \, \mathrm{GeV}$ & 5 & $0.0178626$ & $1.64668$ \\
\hline
\end{tabular}
\caption{3D EQCD parameters for four typical scenarios.}
\label{match_scenarios}
\end{table}

The determination of $\ycrit$ faces the usual challenges of
supercritical slowing down, associated with determining a first order
phase transition point numerically.  In the next section we present a
methodology for evading this problem.

\section{Our method}
\label{sec:method}

The standard way of determining $\ycrit$ would be by applying
multicanonical reweighting in order to enforce tunneling between the
two phases \cite{Kajantie:1998yc}.  This is rather inefficient, so we
develop another approach to efficiently determine the transition
temperature of a first-order phase transition on the lattice.

The main idea is to prepare a lattice configuration where the two
phases coexist and are permanently compared to each other at the
phase boundaries. If we miraculously guessed the exact value of
$y_{\mathrm{crit}}$, the symmetric phase volume would change only via
Brownian motion.  If our value for $y$ were close to but not exactly
$\ycrit$, the phase boundaries would feel a small net pressure, and would
tend to allow the preferred phase to expand at the expense of the
other.  This leaves us with two questions:
\begin{itemize}
\item How do we prepare such configurations?
\item How can we tune the mass to its critical value and balance the
  Brownian motion of the phase boundaries?
\end{itemize} 

\begin{figure}[htbp!] 
\centering
	\subfigure[$\frac{1}{N_x N_y} \sum_{x,y} \Tr \, \left( \Phi_{\mathrm{L}}^3(\bx) \right)$]{\includegraphics[scale=0.5]{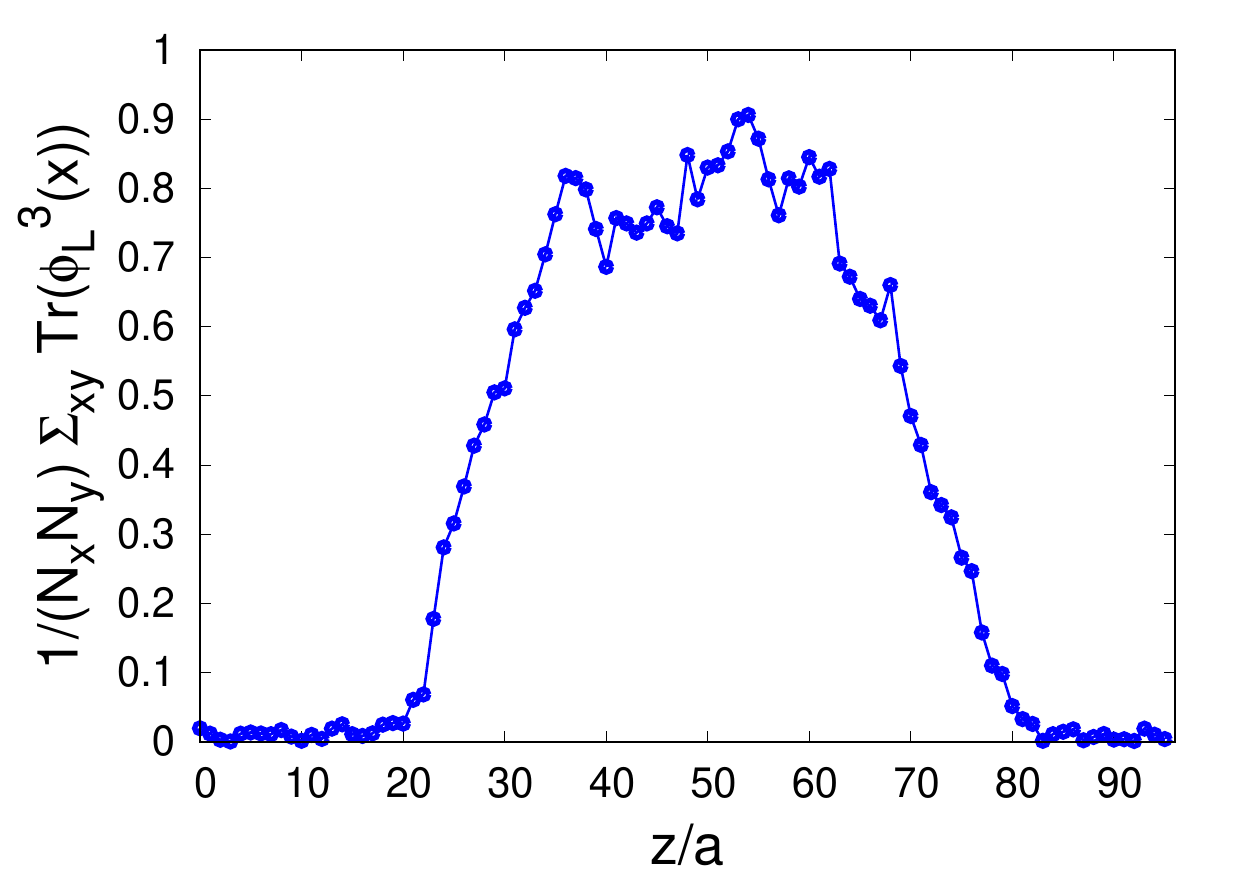}} 
	\subfigure[$\frac{1}{N_x N_y} \sum_{x,y} \Tr \, \left( \Phi_{\mathrm{L}}^2(\bx) \right)$]{\includegraphics[scale=0.5]{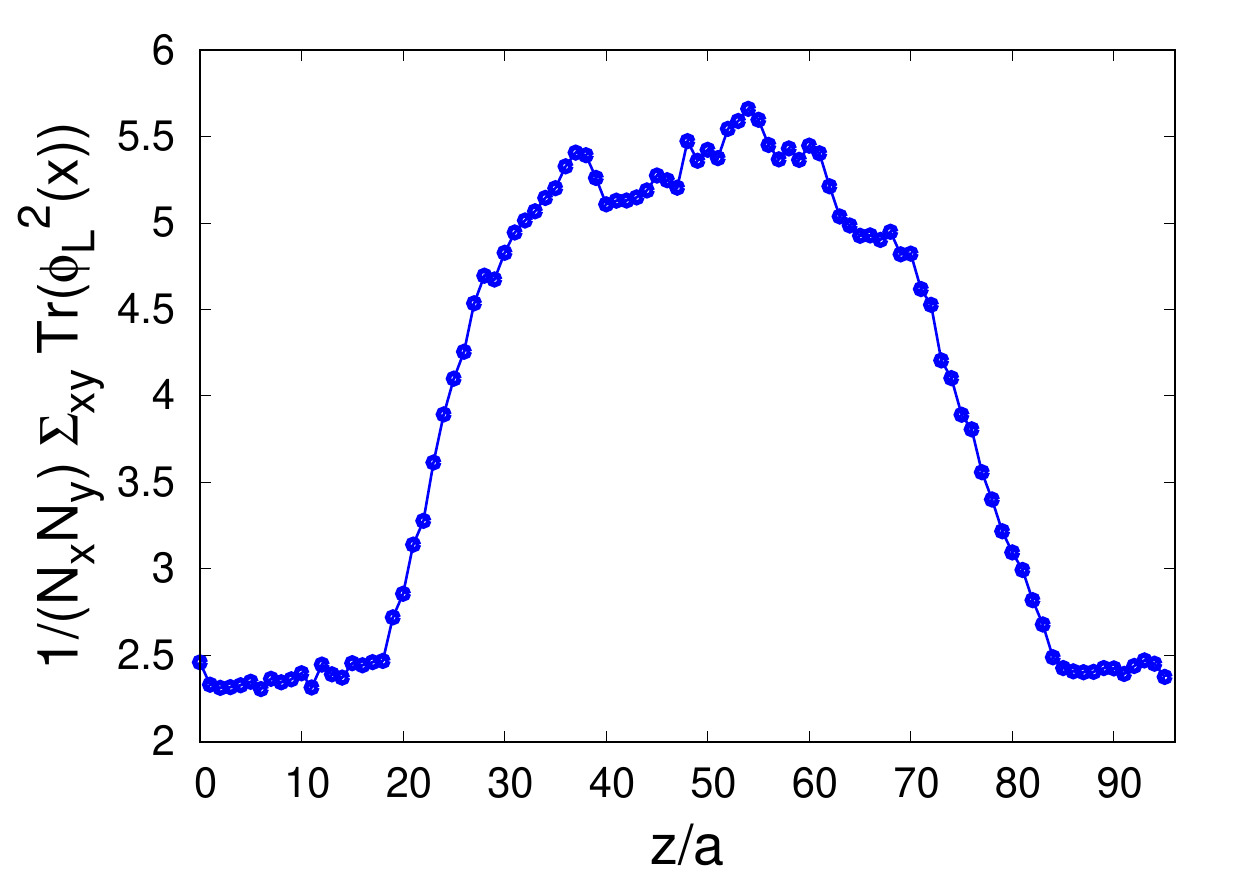}} 
	\caption{$\Tr\, \Phi^3$ and $\Tr \, \Phi^2$, integrated over
          transverse directions, as a function of the $z$ direction in
          a $36^3 \times 96$ box at $\gsqa=\frac 13$, $x=0.08896$
          and $y=0.47232$.
          There is a region near 0 (periodically identified with 96)
          which is in the symmetric phase, and a region near 50 which
          is in the broken phase, as well as two phase boundaries.
          The phases are visible in either order parameter but the
          fluctuations in $\Tr \, \Phi^2$ are smaller.}
	\label{slices}
\end{figure} 

The true order parameter of EQCD is
$\Tr \, \Phi^3 $, which indicates whether the
$\mathbb{Z}_3$-symmetry of $\Phi$ is present or
broken. However, the phase transition can also be spotted in
$\Tr \, \Phi^2 $ (see Fig.~\ref{slices}),
which has smaller fluctuations and leads to a more stable phase
discriminator; so we use it in the following.  Our approach begins by
bounding $\ycrit$ by performing a simulation in a modest-sized cubic
box, starting from a quite positive $y$ value and decreasing it after each
update sweep.  At some value, $\Tr\, \Phi^2$ abruptly jumps.  Then one
steadily increases $y$ until the value abruptly falls.  This
determines upper and lower \textsl{spinodal} $y$-values; $\ycrit$ must
lie between, typically close to the upper value.

Next we estimate $\Tr \, \Phi^2_{\mathrm{symm}}$ and
$\Tr \, \Phi^2_{\mathrm{brok}}$, the values of $\Tr \, \Phi^2_{\mathrm{L}}$
in each of these phases at a mass close to the transition temperature,
which we do in separate simulations which are initialized with either
vanishing or large constant $\Phi$ values.  The method will be rather
insensitive to the exact values of these quantities, so it is not
important if the determinations are from somewhat incorrect $y$
values.

Next we set up our mass tuning
algorithm.  We work in a rectangular periodic lattice with one long
($L_z$) direction and two equal shorter ($L_x=L_y$) directions.
Initially we make the $y$ (Lagrangian) value $z$-coordinate dependent,
\begin{equation}
  \label{zdependenty}
  y(z) = y_{\mathrm{crit,est}} + \Delta y \cos(2\pi z/L_z) \,,
\end{equation}
with $\Delta y$ chosen initially to be large enough that
$y_{\mathrm{crit,est}} \pm \Delta y$ are above/below the spinodal
values.  After a series of update sweeps, the field will
find the symmetric phase where $y$ is large and the broken phase where
$y$ is small, generating our configuration with both phases and two
phase boundaries.  Then the magnitude of $\Delta y$ is gradually lowered
over a series of update sweeps; if one phase starts to win out over
the other, the estimated critical value is adjusted.

With our starting two-phase configuration and estimated $\ycrit$ in
hand, we proceed to the more accurate determination of $\ycrit$.  We
continue to evolve with a space-uniform $y$ value, but we adjust it
after each lattice site-update according to
\begin{equation}
\label{mass_adjustment}
y_{\mathrm{L,\,new}} = y_{\mathrm{L,\,old}} + \cb \cdot \left(
\frac{\tfrac{1}{V} \sum_\bx \Tr \, \Phi^2 - \Tr \,
  \Phi^2_{\mathrm{symm}}}{\Tr \, \Phi^2_{\mathrm{brok}} - \Tr \, 
  \Phi^2_{\mathrm{symm}}}  - 0.5 \right) \, ,
\end{equation}
where $y_{L} \equiv Z_3(y + \delta y_{2\mathrm{loop}})$ and $\cb$ is a
small coefficient that controls the strength of the adjustment.
The quantity in brackets here is an estimate for the fraction of the
volume which lies in the broken phase, based on the known
(approximate) values of $\Tr \, \Phi^2$ in each phase.  Therefore, the
adjustment term shifts $y$ upwards (making the symmetric phase more
preferred) when more volume is in the broken phase, and downwards
(making the broken phase more preferred) if more of the volume is
symmetric.

The coefficient $\cb$ is small,
$\OO \left( \frac{1}{N_\mathrm{x} N_\mathrm{y} N_\mathrm{z}} \right)$, such that
the evolution of $y$ is as mild as possible, consistent with enough
restorative force to prevent either phase from ``winning.''
Specifically, whenever $y$ deviates from $\ycrit$, there is a net
force on the interface, equal to the surface area times $\Delta F$ the
free energy difference between phases.  At $y=\ycrit$, $\Delta F=0$
and there is no net force on the interface.  Away from $y=\ycrit$, we
can expand $\Delta F$ in a Taylor series in $y-\ycrit$.  At leading
order in small $y-\ycrit$,
the free energy difference will be linear in $y-\ycrit$,
and the central value of $y$ which maintains coexistence will equal
$\ycrit$.
At quadratic order, $d^2 F/dy^2 \neq 0$ means that the
restorative force is slightly biased, and we will obtain an incorrect
value for $\ycrit$.  We test for such a distortion by performing a
second evolution where $\cb$ is twice as large, to confirm that the
central value of $y$ is the same within errors (which it is in all
cases we considered).

\section{Results}
\label{sec:results}

We use the procedure described in the previous section to determine
the critical value $\ycrit(x,a)$ for several values of the scalar
self-coupling $x$, each at several lattice spacings.  The exact list
of lattices considered is given in Table \ref{EQCD_sim_params}.
Because our procedure leads to relatively long autocorrelation in the
estimated $\ycrit$ value, the errors must be determined via the
jackknife method using relatively wide jackknife bins; we vary the bin
widths until the error estimates stabilize.
We then
subtract the known 1- and 2-loop contributions and apply the known
multiplicative rescalings \cite{Moore:1997np}
from the results and convert
$y_\mathrm{L,\, crit} \rightarrow y_{2\mathrm{loop} ,\, \mathrm{crit}}$.
For each $x$ value, we must extrapolate this quantity to zero lattice
spacing; the intercept is the continuum critical $y$ value and the
slope at the intercept is the desired $\OO(a)$ additive correction to
the scalar mass.

\begin{figure}[htbp!] 
\centering 
	\begin{tabular}{cc}
		\includegraphics[scale=0.5]{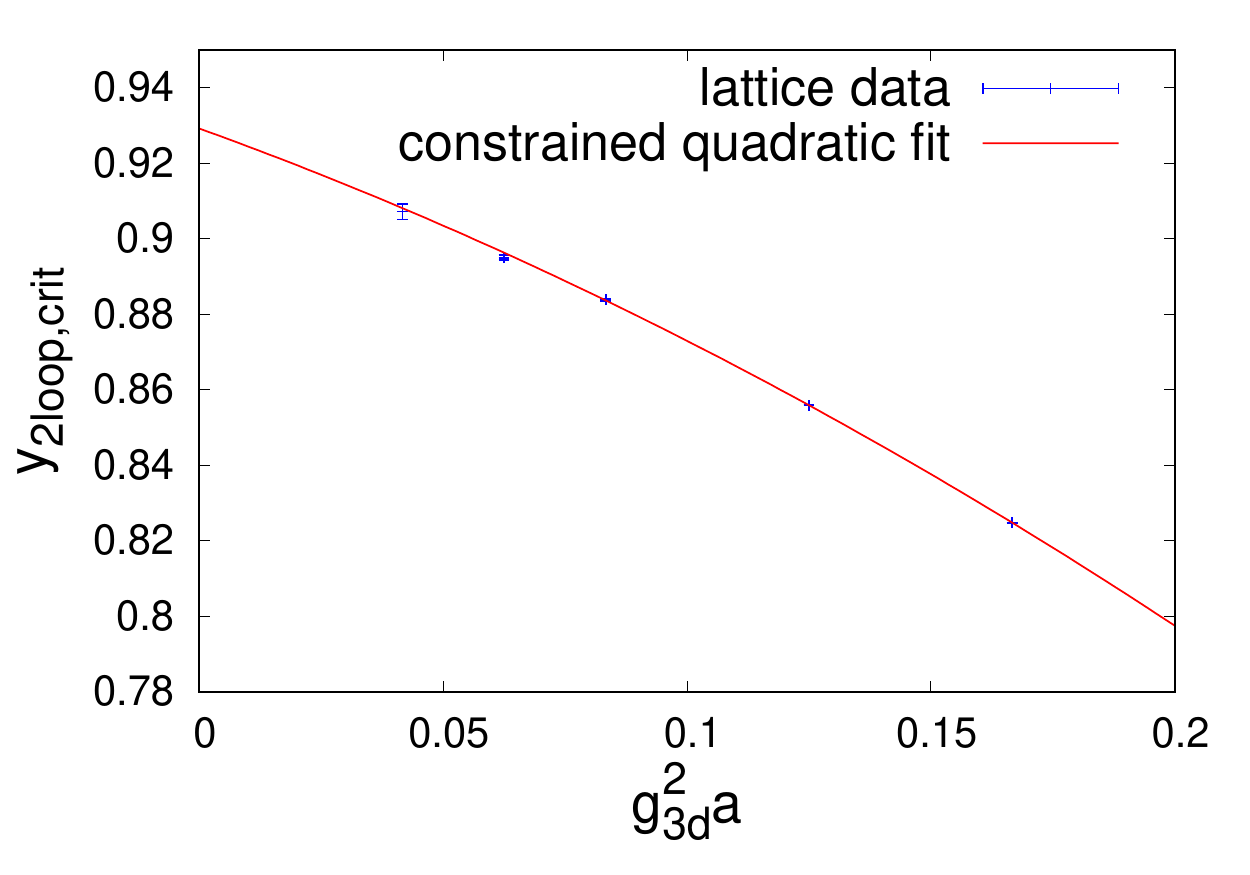} & 
		\includegraphics[scale=0.5]{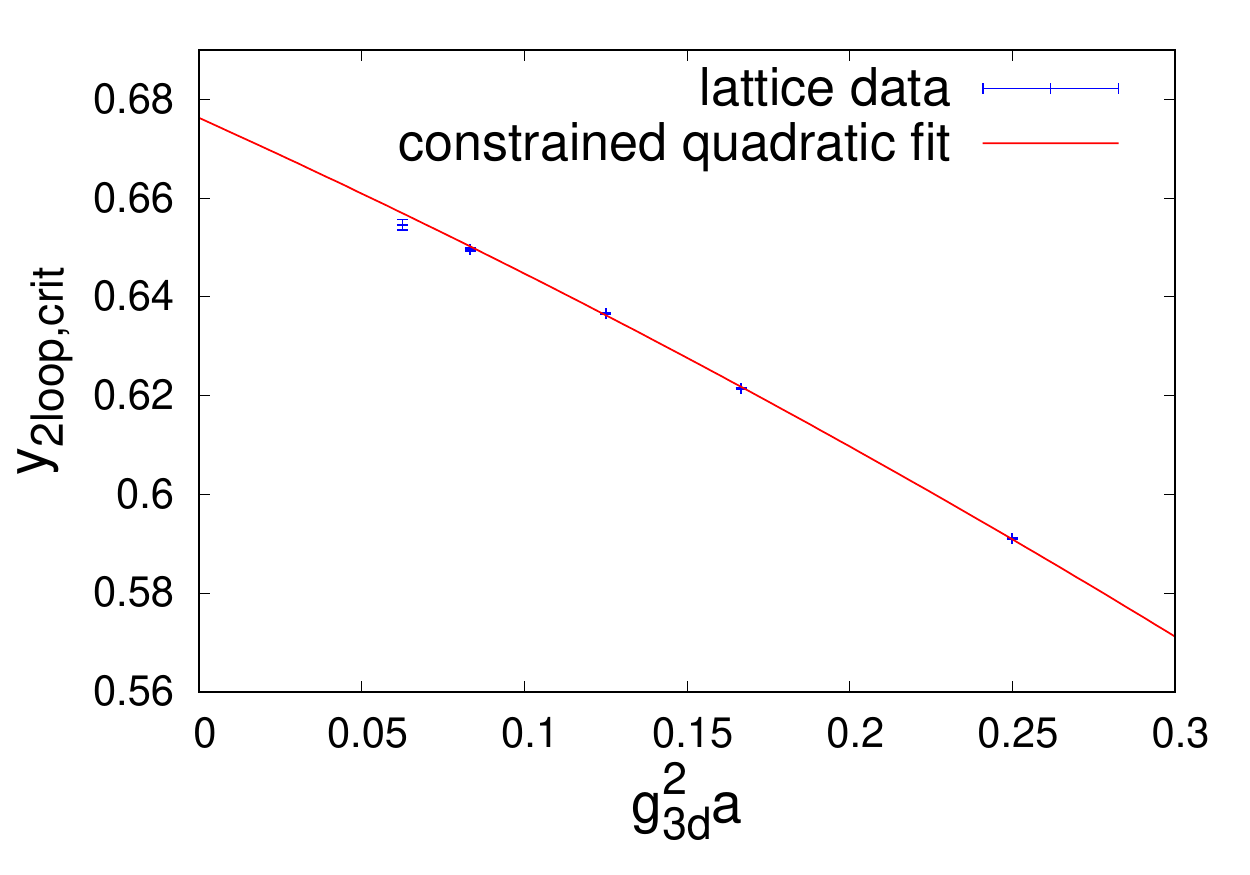} \\
		(a) $x=0.0463597$ & (b) $x=0.0677528$ \\
		\includegraphics[scale=0.5]{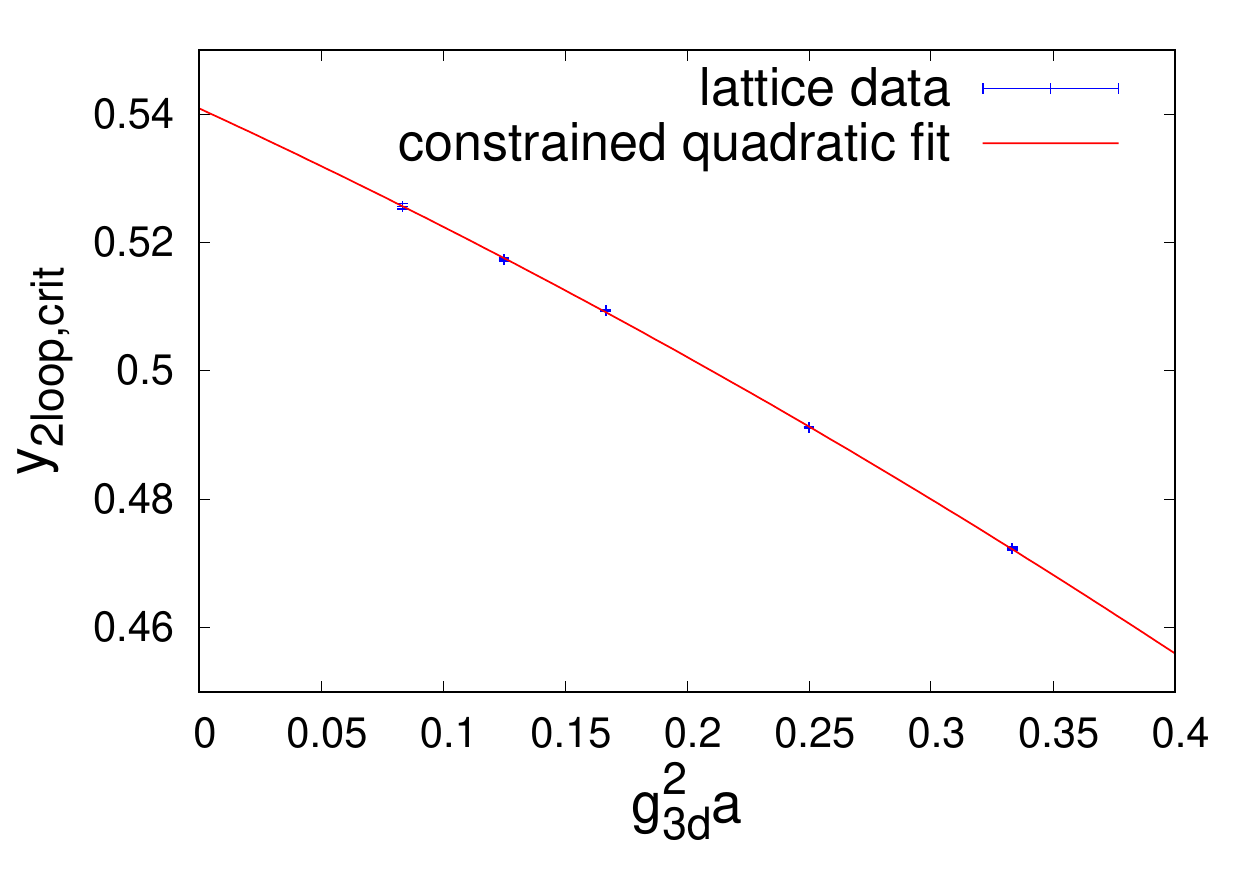} &
		\includegraphics[scale=0.5]{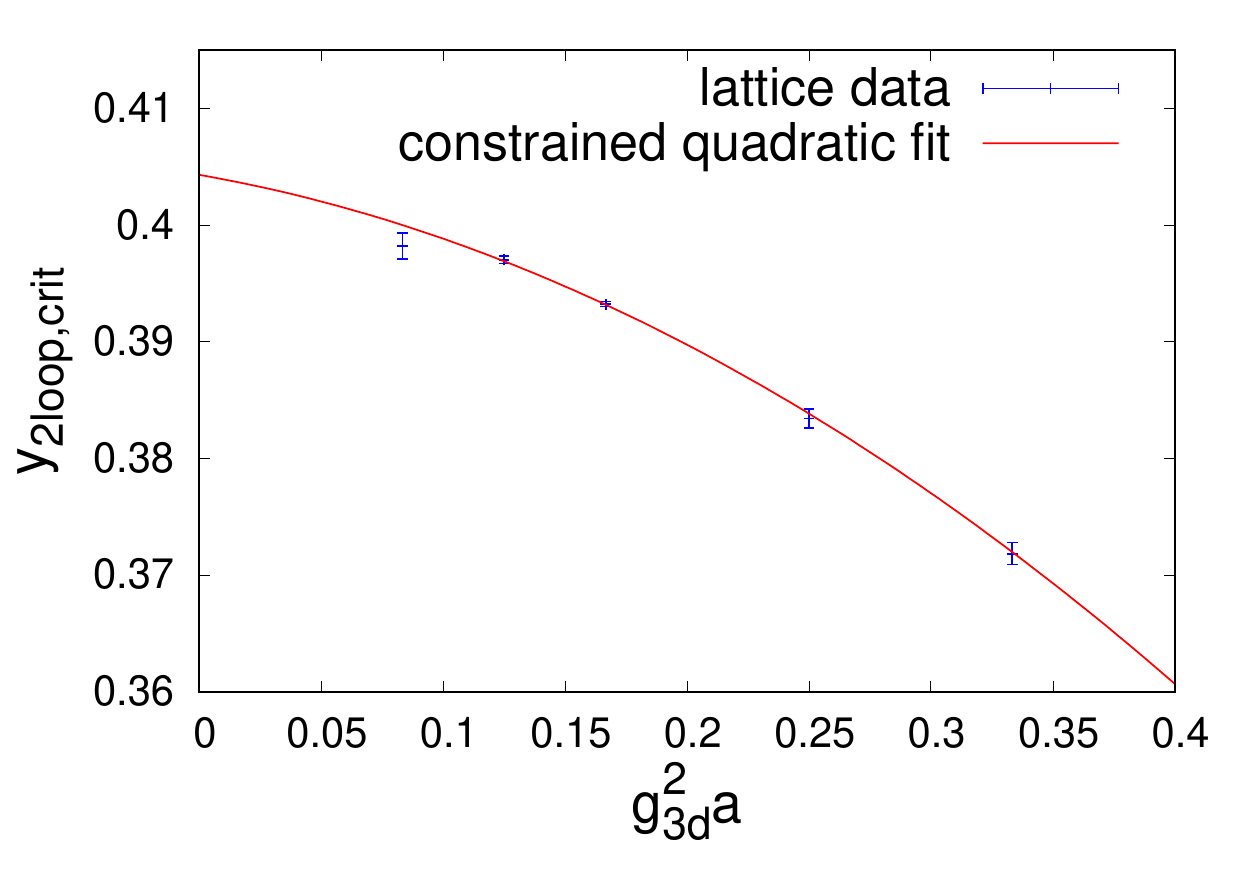} \\ 
		(c) $x=0.08896$ & (d) $x=0.13$ \\
		\includegraphics[scale=0.5]{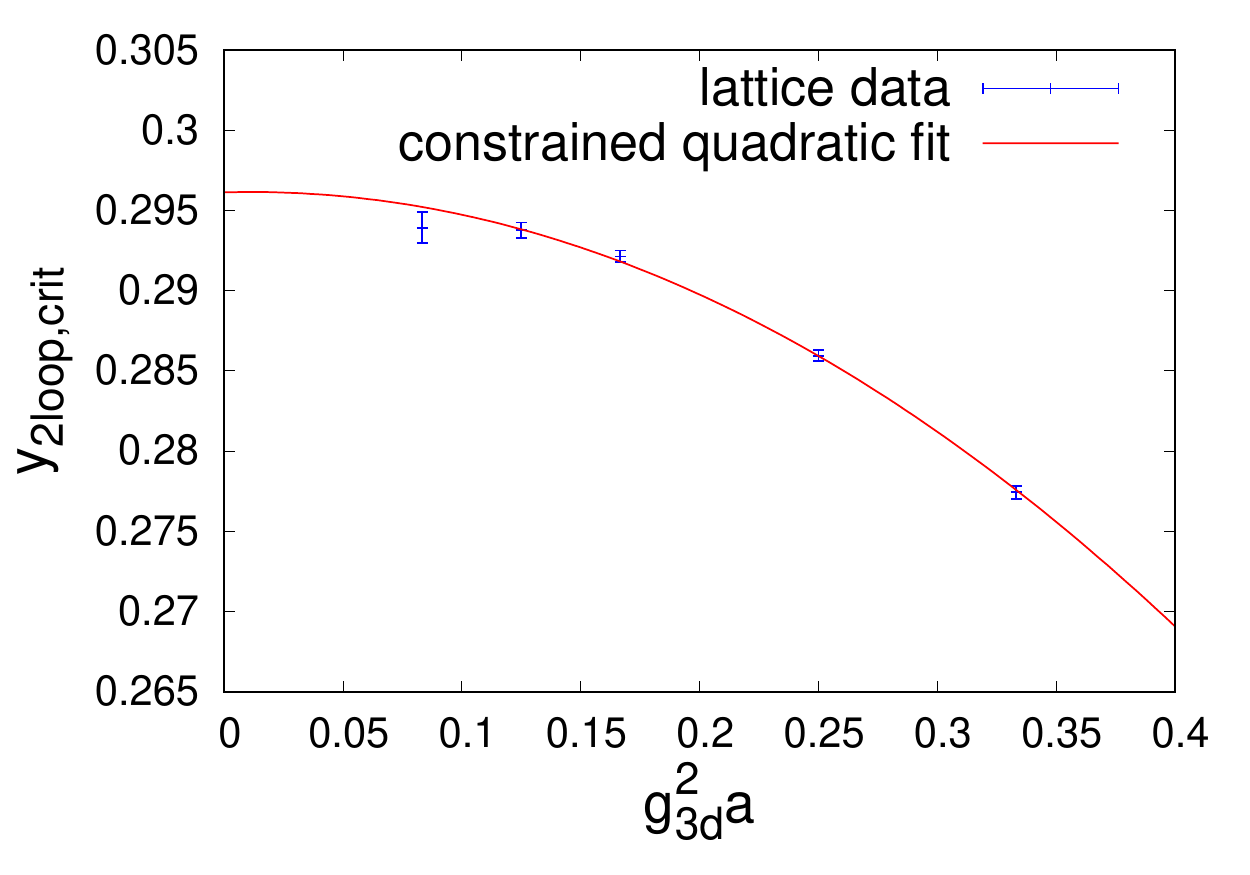} &
		\includegraphics[scale=0.5]{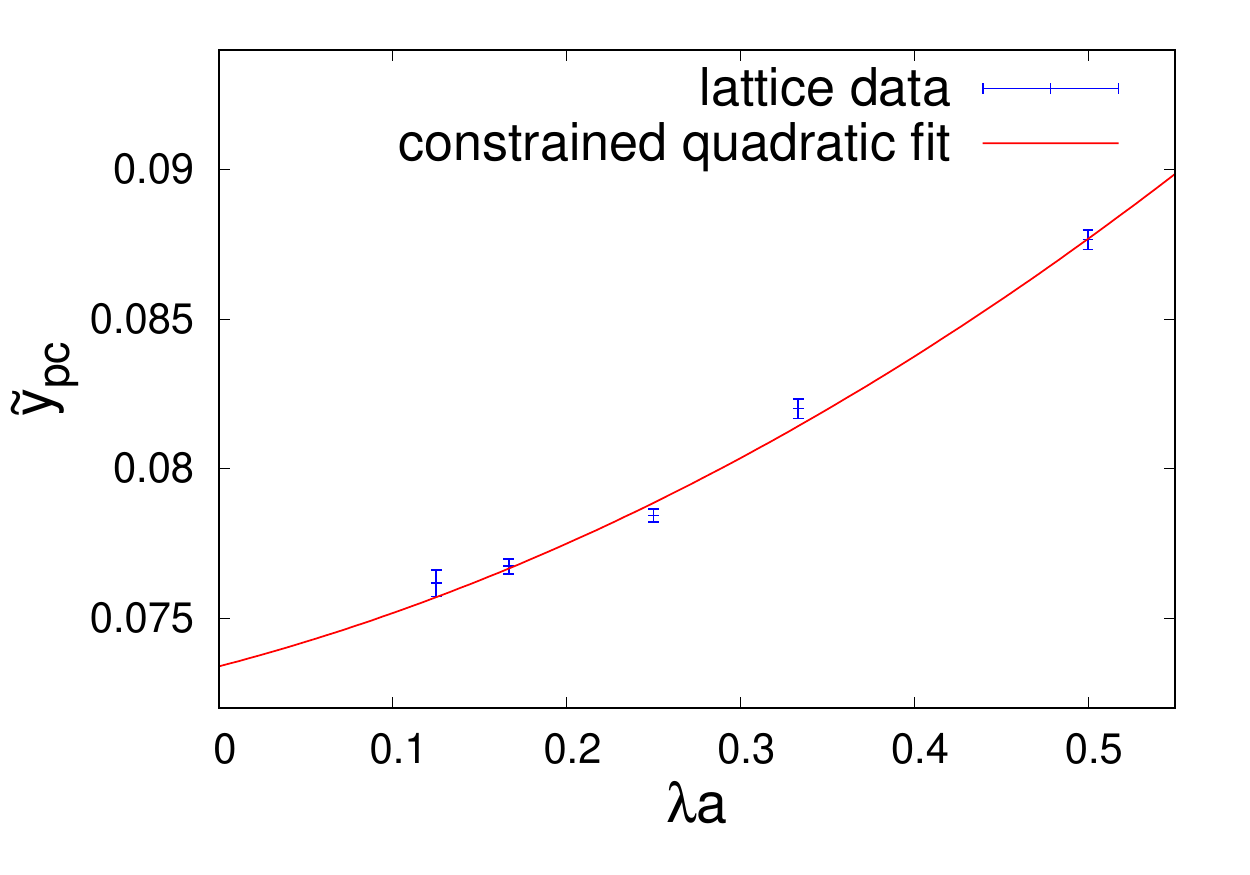} \\ 
		(e) $x=0.2$ & (f) pure scalar 
	\end{tabular}
	\caption{Fits of $\OO(\gsqa)$ behavior for different $x$.}
	\label{fits_data}
\end{figure}

Because our $\ycrit(x,a)$ results are quite precise but the $a$ values
are not extremely small, we anticipate that $\ycrit(x,a)$ contains
corrections beyond linear order in $a$.
In principle, we could straightforwardly fit a polynomial of order
$N_\mathrm{poly}$ in $\gsqa$ as
\begin{equation}
  \label{simplefit}
\ycrit \left( \gsqa \right) = \sum_{j=0}^{N_{\mathrm{poly}}} \, y_j (\gsqa)^j \, .
\end{equation}
However, as often occurs, ever-higher order coefficients are ever less
certain, and including too many coefficients tends to overfit the data
and artificially inflates the final fitting errors.
In order to extract these coefficients as efficiently as possible from
the data, we would like to build in our knowledge about the
convergence of the perturbative series to the fit. A useful tool to
implement this is constrained curve fitting \cite{Davies:1994mp,Lepage:2001ym}. 
Motivated by a rough estimate of the radius of convergence
$(\gsqa)_{\mathrm{conv}} \approx 0.5$, we make the a priori-guess
\begin{equation}
\label{scalingest}
  | y_i | \leq \frac{y_0}{2^i} \, ,
\end{equation} 
having obtained $y_0$ from a standard, unconstrained fit.  We then
use this estimate to choose the size of a zero-centered chisquare
prior on each fitting parameter.
The procedure has almost no impact on the determined values of $y_0$
and $y_1$, where the data is far more constraining than the prior.
In practice, a quadratic polynomial is sufficient to give a good fit
with a reasonable $\chi^2$.
The results of these fits are given in Table \ref{res_coeffs} and the
fits themselves are displayed in Fig.~\ref{fits_data}.
We also confirmed by varying the volume that any finite-volume effects
are smaller than our statistical error bars.

\begin{table}[htbp!]	
\centering
\begin{tabular}{|c|c|c|}	
\hline
$x$ & $y_\mathrm{crit,\,cont}$ & $\delta y_{3\mathrm{loop}} / \gsqa $  \\
\hline
$0.0463596$ & $0.9293(13)$ & $-0.467(19)$ \\
$0.0677528$ & $0.67627(85)$ & $-0.298(10)$ \\
$0.08896$ & $0.54092(76)$ & $-0.1750(74)$ \\
$0.13$ & $0.4043(18)$ & $-0.037(18)$ \\
$0.2$ & $0.2961(15)$ & $0.004(15)$ \\
\hline
\end{tabular}
\caption{Results of our five EQCD simulation sets.}
\label{res_coeffs}
\end{table}

We caution the reader that, while $y_0$ and $y_1$ can be interpreted
as the continuum critical point and the 3-loop $\OO(a)$ additive mass
renormalization coefficient, we cannot interpret $y_2$ as a 4-loop
mass renormalization or use it to further improve the
lattice-continuum matching.  That is because there are many
uncontrolled $\OO(a^2)$ corrections which influence $y_2$.  For
instance, there are unknown 2-loop $\OO(a^2)$ lattice-continuum
corrections to $x$, which influence the critical value via
$(d\ycrit/dx) \delta x$.  Similarly, tree-level $\OO(a^2)$
high-dimension operators and $\OO(a^2)$ corrections to $g^2$ (which we
could interpret as uncertainties in the scale setting) also lead to
$\OO(a^2)$ effects in the $\ycrit$ value.
Because $(d\ycrit/dx) \sim x^{-2}$, the $\OO(a^2)$ and other
higher-order effects will become severe as we go towards small $x$
values.  Therefore small $x$ requires the use of very fine lattices.
Furthermore, when $x$ is small, there becomes a hierarchy of mass
scales in the problem;
$m_{A,\mathrm{brok}} \gg m_{\Phi} \gg m_{A,\mathrm{symm}}$.  Both
effects make the accurate extraction of $y_1$ at small $x$ very
numerically demanding.  Therefore we did not treat the smallest $x$
value shown in Table \ref{match_scenarios}.  Instead, we add two
larger values of $x$, $x=0.13$ and $x=0.20$, which are still within
the domain where the transition is first order, but which give us a
broader $x$ range over which to fit $y_1$ as a function of $x$.

\begin{figure}[htbp!]  
\centering
	\includegraphics{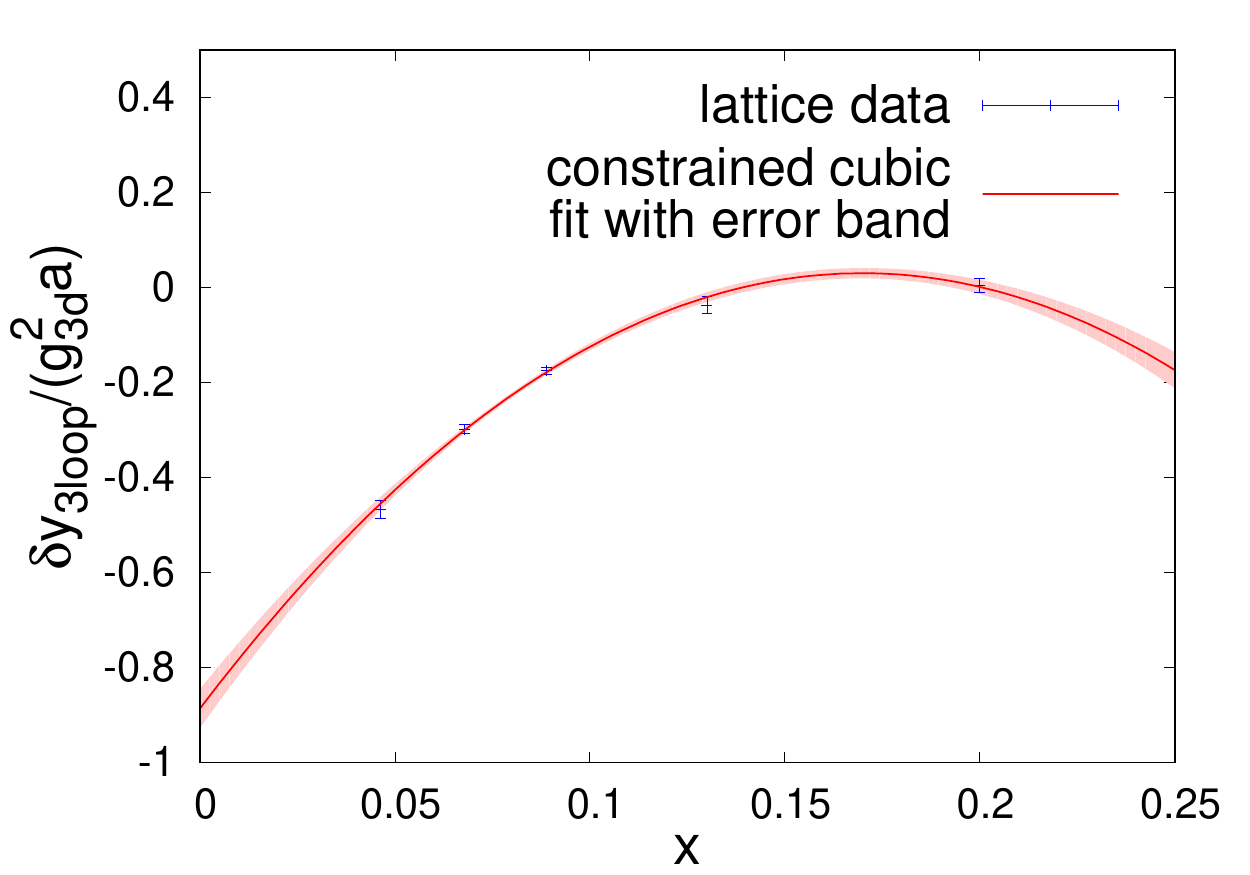} 
	\caption{Grand fit of $\frac{\delta y_\mathrm{3loop}}{\gsqa}(x)$ with error band.}
	\label{grand_fit}
\end{figure}

Next we use our results for $y_1(x)$ to fit its overall $x$ dependence.
The parametric form of the $\OO(\gsqa)$-correction
\cite{DOnofrio:2014mld} was given in \eqref{parametric}.
The $x^3$ coefficient corresponds to 3-loop diagrams containing only
scalar lines.  It is therefore equal to the $\OO(a)$ mass
renormalization term in the theory in the $\gsq \to 0$ limit, which is a
scalar theory.  We explain our (different) procedure to treat this
scalar theory in App.~\ref{app1}; our analysis leads to the result
\begin{equation}
\label{C3val}
  C_3 = \frac{\delta \tilde{y}_\mathrm{3loop}}{\lambda a} = 0.0151(55) \,.
\end{equation}
Here $\tilde y \equiv m^2(\mu=\lambda)/\lambda^2$ is the scalar mass,
made dimensionless using the scale $\lambda$ rather than the scale
$\gsq$; it equals $y/x^2$ up to the effect of the different
renormalization scale.
We incorporate this result as a prior in fitting a cubic polynomial to
the results of Table \ref{res_coeffs}.  The resulting fit,
\begin{equation}
  \label{mainresult}
\frac{\delta y_{3\mathrm{loop}}}{\gsqa}(x) = 0.0151(55) \,
x^3 - 31.8(28) \, x^2 + 10.80(74) \, x - 0.886(41) \, ,
\end{equation}
is displayed in Fig.~\ref{grand_fit}.  We report the full error
covariance matrix in Table \ref{cov_mat_grand_fit}.
This fit constitutes our main result.

\begin{table}
\centering
\begin{tabular}{|c|c c c c| }	
\hline
$\mathrm{cov}(C_i,C_j)$ & $C_0$ & $C_1$ & $C_2$ & $C_3$   \\
\hline
$C_0$ & $0.001700$ & $-0.02997$ & $0.1101$ & $-3.563 \cdot 10^{-8}$ \\
$C_1$ & $-0.02997$ & $0.5451$ & $-2.046$ & $1.129 \cdot 10^{-6}$ \\
$C_2$ & $0.1101$ & $-2.046$ & $7.899$ & $-1.079 \cdot 10^{-5}$ \\ 
$C_3$ & $-3.563 \cdot 10^{-8}$ & $1.129 \cdot 10^{-6}$ & $-1.079 \cdot 10^{-5}$ & $3.025 \cdot 10^{-5}$ \\
\hline
\end{tabular}
\caption{Covariance matrix of the grand fit.}
\label{cov_mat_grand_fit}
\end{table}

\begin{figure}[htbp!] 
\centering 
	\includegraphics{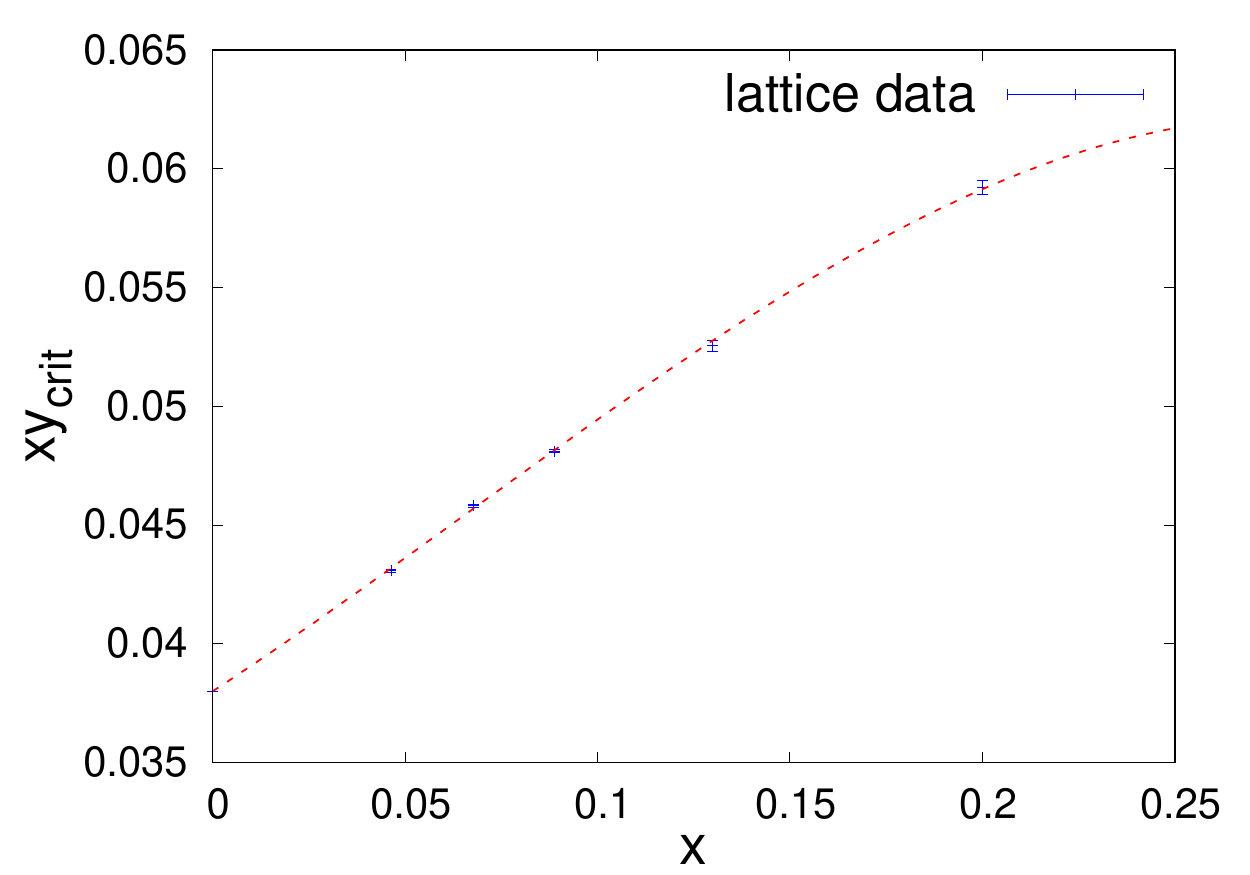} 
	\caption{Updated version of the phase diagram of EQCD. The
          phase below the data points is the $\mathbb{Z}_3$-broken
          phase, the one above is $\mathbb{Z}_3$-symmetric.
          The dashed line is a fit to the datapoints, to help guide
          the eye.}
	\label{xy_plot}
\end{figure}

As a corollary, we provide an updated version of the EQCD phase
diagram. The version from \cite{Kajantie:1998yc} does not include
continuum-extrapolated critical masses. The intercept of our EQCD
fits delivers these critical masses. Additionally, the
$x \rightarrow 0$ limit
\begin{equation}
  \label{xycritlim}
x\ycrit = \frac{3}{8 \pi^2}
\end{equation}
is known perturbatively \cite{Kajantie:1998yc}.
We present our data, and this limiting value, in Fig.~\ref{xy_plot}.
In addition, to guide the eye%
\footnote{In fact we
  expect nonanalytical behavior as $x\to 0$, due for instance to the
  two-loop $\Phi^2 \ln(\Phi^2/\mu)$ terms in the effective potential
  \cite{Arnold:1992rz} which give rise to
  $x\ycrit - 3/8\pi^2 \sim x \ln(x)$ corrections to
  Eq.~(\ref{xycritlim}).},
we include a cubic fit of $x\ycrit$ as a function of $x$, displayed by a dashed 
line. There is quantitative agreement with
the phase diagram in \cite{Kajantie:1998yc} at small $x$, but at large
$x$ we find that the prominent bending down of the $x\ycrit$ curve
found by \cite{Kajantie:1998yc} arose because they failed to take a
continuum limit.

Kajantie \textsl{et al} \cite{Kajantie:1998yc} found that the
tricritical point occurs at $x=0.25$, beyond which the phase
transition becomes  of second order.
We have not studied $x$ values larger than $x=0.2$, so we cannot make
any statement about the location of the tricritical point.

As a byproduct our study also produces values for the discontinuity in
$\Tr \, \Phi^2$ across the phase transition point, and for the
$\OO(a)$ additive correction to the $\Tr \, \Phi^2$ operator, which
was also not previously known.  We postpone these secondary results to Appendix \ref{app2}.

\section{Conclusion and outlook}
\label{sec:conclusion}

We have computed the remaining $\OO(a)$ improvement coefficient
in the lattice-continuum matching of 3D EQCD (SU(3) gauge theory with
an adjoint scalar in 3 space dimensions).  We did so by using the
first order phase transition point as a fixed physical point.
We developed a new methodology to efficiently extract the critical
scalar mass where the first order transition occurs.
Determining the critical scalar mass $\ycrit(x,a)$ as a function of
$a$ for fixed $x$ allows an extrapolation to the continuum limit; the
linear term in the extrapolation is the desired improvement
coefficient.  We then performed a grand fit to its known functional
form.  As a byproduct, a continuum-extrapolated update of the
EQCD phase diagram was obtained.

Now that we are in possession of the last missing ingredient, we aim to
compute the modified EQCD-Wilson-loop
that leads to $C(q_\perp)$ and extrapolate it to continuum. The
continuum extrapolation is drastically facilitated by our completion
of the renormalization \cite{Panero:2013pla}. The jet broadening
coefficient $\hat{q}$ can be derived as the second moment of
$C(q_\perp)$.  We hope that the resulting nonperturbative information
on $C(q_\perp)$ will be of utility in studying and interpreting jet
modification arising from the hot QCD medium in relativistic heavy ion
collisions.

\section*{Acknowledgments}
This work was supported by the Deutsche Forschungsgemeinschaft (DFG,
German Research Foundation) – project number 315477589 – TRR
211. Calculations for this research were conducted on the Lichtenberg
high performance computer of the TU Darmstadt.  We thank Kari
Rummukainen and Aleksi Kurkela for useful conversations, and Daniel
Robaina for his patient help with the OpenQCD-1.6 codebase.

\appendix
\section{Algorithm for pure scalar case}
\label{app1}

The large-$x$ limit of EQCD is the same as the $\gsq \to 0$ limit
(provided we work in terms of
$\tilde y \equiv y/x^2 = m^2/\lambda^2$ and track the lattice spacing
in terms of $\lambda a = \gsqa\, x$).
In this limit, we have an 8 (real) component scalar theory with an $\OO(8)$
symmetry and a second-order phase transition where this symmetry is
spontaneously broken to $\OO(7)$.  The $x^3$ term in the scalar mass
renormalization of EQCD arises purely from scalar diagrams which are
identical to those in this theory; therefore we can determine this
coefficient by studying the $\OO(\lambda a)$ corrections to
$\tilde y$ in this scalar field theory.

A natural approach
would be to use, as a line of constant physics, the 2'nd order phase
transition point.  However this would face the usual problems of
critical slowing down and the inaccuracy of establishing the exact
transition point.  So we choose instead to compare $\tilde y$ between
different lattice spacings by finding the \textsl{pseudo}critical
value where the theory in a specific physical volume encounters a
specific pseudocritical criterion.  We choose the volume to be
$\lambda L = 8$ and select as the pseudocritical condition that
the 4th order Binder cumulant \cite{Binder:1981sa}
\begin{equation}
\mathcal{B} \equiv \frac{\langle \left( \Tr \, \bar{\Phi}^2 \right)^2 \rangle}
{\langle \Tr \, \bar{\Phi}^2 \rangle^2} \, ,
\end{equation}
where
$\bar{\Phi} \equiv \tfrac{1}{N^3} \sum_\bx \Phi_\mathrm{L}(\bx)$,
takes the value $\mathcal{B}_{\mathrm{pc}}=1.073$.
Because the Binder cumulant is dominated by infrared physics and is
insensitive to the lattice spacing up to subleading corrections and
renormalization effects (which are what we want to study),
this should occur at the same \textsl{physical} $y$ value at every
lattice spacing up to $\OO(a^2)$ or higher corrections.
Our set of simulation parameters can be found in
Table \ref{scalar_sim_params}.  The results already appeared in the
last panel of Fig.~\ref{fits_data}.

\begin{table}[htbp!] 
\centering
\begin{tabular}{|c|c|c|}	
\hline
$\lambda a$ & $N_\mathrm{x} N_\mathrm{y} N_\mathrm{z}$  & total statistics \\
\hline
$1/2$ & $16^3$ & $402900$ \\ 
$1/3$ & $24^3$ & $1490120$ \\ 
$1/4$ & $32^3$ & $2487440$ \\ 
$1/6$ & $48^3$ & $2931750$ \\ 
$1/8$ & $64^3$ & $2513280$ \\
\hline
\end{tabular}
\caption{Simulation parameters for the pure scalar simulation.}
\label{scalar_sim_params}
\end{table}


\section{EQCD simulation parameters and phase transition properties}
\label{app2}

\subsection{Parameters and table}

In Table \ref{EQCD_sim_params}, we provide parameters and direct results of our 
EQCD simulations as well as values for $\Tr \, \Phi^2$ in both phases at 
criticality. We converted $\Tr \, \Phi^2_\mathrm{L,crit} \rightarrow 
\frac{\Tr \, \Phi^2_\mathrm{crit,cont}}{\gsq}$, where we display the latter in the table, via all known contributions up to $\OO(\gsqa)$ in Eq.~(\ref{phi3loop}) \cite{Moore:1997np}. For the sake of readability, we will refer to $\frac{\Tr \, \Phi^2_\mathrm{crit,cont}}{\gsq}$ as $\phisqcont$ in the following. The raw data was obtained in separate simulations with $V=N_\mathrm{x}^3$. We ensured that the Monte-Carlo error of the separate simulations dominates the overall error of $\phisqcont$, not the uncertainty of $y_\mathrm{crit}$. The slightly negative values of $\frac{\Tr \, \Phi^2_\mathrm{symm}}{\gsq}$ for small $x$ are expected and arise because the positive mass-squared at the transition point suppresses IR fluctuations, while the renormalization involves  subtracting off large positive counterterms 
(including the massless, free-theory fluctuations). 
  
\begin{table}[htbp!] 
\centering {\small
\begin{tabular}{|c|c|c|c|c|c|c|}	
\hline
$\gsqa$ & $x_\mathrm{cont}$  & $N_\mathrm{x} N_\mathrm{y} N_\mathrm{z}$ & $y_
\mathrm{crit,cont}$ & statistics $y_\mathrm{crit}$ & \!$\Tr \, \Phi^2_
\mathrm{brok}/\gsq \!$ & $\Tr \, \Phi^2_\mathrm{symm}/\gsq$  \\
\hline
$1/6$ & $0.0463597$ & $72^2 \times 192$ & $0.824773(45)$ & $300690$ & 
$14.0477(40)$ & $-0.22234(28)$ \\
$1/8$ & $0.0463597$ & $96^2 \times 256$ & $0.855935(85)$ & $373030$ & 
$15.4237(89)$ & $-0.23064(43)$ \\
$1/12$ & $0.0463597$ & $144^2 \times 384$ & $0.88387(28)$ & $37170$ & 
$17.716(12)$ & $-0.23390(77)$ \\
$1/16$ & $0.0463597$ & $192^2 \times 512$ & $0.89504(62)$ & $1300$ & 
$17.714(14)$ & $-0.23748(74)$ \\
$1/24$ & $0.0463597$ & $192^2 \times 512$ & $0.9072(21)$ & $1430$ & $17.585(57)
$ & $-0.2355(17)$ \footnote{For the sake of feasibility, we did not scale the 
volume here, accordingly. However, ref. \cite{Hietanen:2008tv} states that no 
finite volume effects occur for $\beta = \tfrac{6}{\gsqa}$ being smaller than 
the smallest extend of the lattice.} \\
\hline
$1/4$ & $0.0677528$ & $48^2 \times 128$ & $0.59100(19)$ & $13110$ & $6.7502(59)
$ & $-0.11392(62)$ \\
$1/6$ & $0.0677528$ & $72^2 \times 192$ & $0.62149(12)$ & $150000$ & 
$7.6155(72)$ & $-0.12043(55)$ \\
$1/8$ & $0.0677528$ & $96^2 \times 256$ & $0.63661(11)$ & $200000$ & $7.960(14)
$ & $-0.12101(78)$ \\
$1/12$ & $0.0677528$ & $144^2 \times 384$ & $0.64963(29)$ & $38990$ & 
$8.287(15)$ & $-0.12400(80)$ \\
$1/16$ & $0.0677528$ & $192^2 \times 512$ & $0.6546(11)$ & $1910$ & $8.4103(61)
$ & $-0.1228(12)$ \\
\hline
$1/3$ & $0.08896$ & $36^2 \times 96$ & $0.47232(24)$ & $10000$ & $4.0728(22)$ & 
$-0.03277(47)$ \\
$1/4$ & $0.08896$ & $48^2 \times 128$ & $0.49119(14)$ & $100000$ & $4.4343(53)$ 
& $-0.03489(59)$ \\
$1/6$ & $0.08896$ & $72^2 \times 192$ & $0.50940(15)$ & $150000$ & $4.7728(35)$ 
& $-0.0344(10)$ \\
$1/8$ & $0.08896$ & $96^2 \times 256$ & $0.51736(20)$ & $97500$ & $4.9249(58)$ 
& $-0.0336(11)$ \\
$1/12$ & $0.08896$ & $144^2 \times 384$ & $0.52565(45)$ & $44530$ & $5.052(10)
$ & $-0.0321(13)$ \\
\hline
$1/3$ & $0.13$ & $36^2 \times 96$ & $0.37184(94)$ & $10000$ & $2.3337(64)$ & 
$0.1112(15)$ \\
$1/4$ & $0.13$ & $48^2 \times 128$ & $0.38341(83)$ & $100000$ & $2.3899(64)$ & 
$0.1126(19)$ \\
$1/6$ & $0.13$ & $72^2 \times 192$ & $0.39323(21)$ & $150000$ & $2.4703(75)$ & 
$0.1190(38)$ \\
$1/8$ & $0.13$ & $96^2 \times 256$ & $0.39702(31)$ & $109230$ & $2.5371(82)$ & 
$0.1246(39)$ \\
$1/12$ & $0.13$ & $144^2 \times 384$ & $0.3982(11)$ & $120600$ & $2.618(13)$ & 
$0.1332(65)$ \\
\hline
$1/3$ & $0.2$ & $36^2 \times 96$ & $0.27743(41)$ & $12340$ & $1.3115(74)$ & 
$0.578(32)$ \\ 
$1/4$ & $0.2$ & $48^2 \times 128$ & $0.28595(34)$ & $21030$ & $1.283(13)$ & 
$0.630(34)$ \\ 
$1/6$ & $0.2$ & $72^2 \times 192$ & $0.29214(37)$ & $66160$ & $1.2850(81)$ & 
$0.650(24)$ \\ 
$1/8$ & $0.2$ & $96^2 \times 256$ & $0.29377(48)$ & $122080$ & $1.323(11)$ & 
$0.713(37)$ \\ 
$1/12$ & $0.2$ & $144^2 \times 384$ & $0.29393(97)$ & $3330$ & $1.390(14)$ & 
$0.707(26)$ \\
\hline
\end{tabular} }
\caption{Parameters and results of EQCD simulations. }
\label{EQCD_sim_params}
\end{table}

\subsection{Transition strength}

\begin{figure}[htbp!] 
\centering 
	\begin{tabular}{cc}
		\includegraphics[scale=0.5]{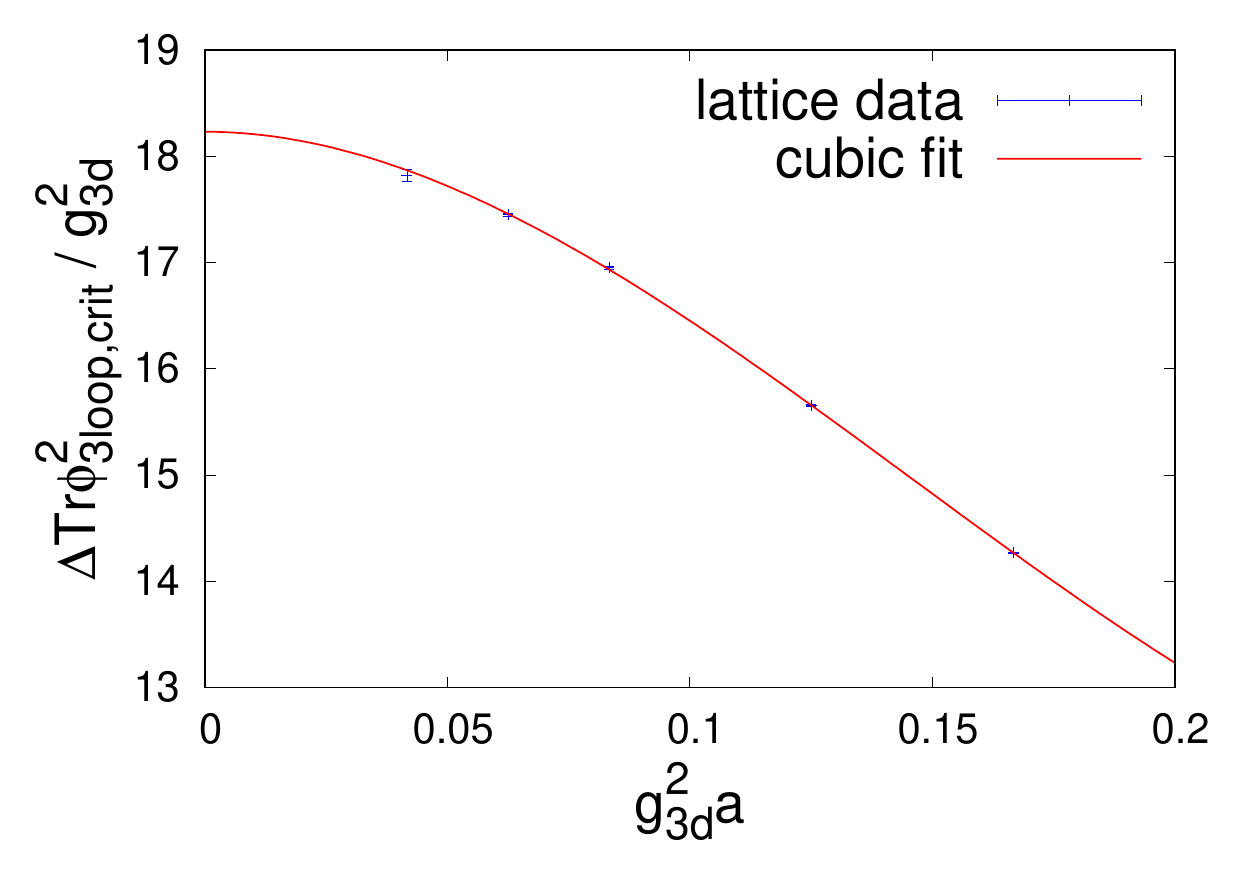} & 
		\includegraphics[scale=0.5]{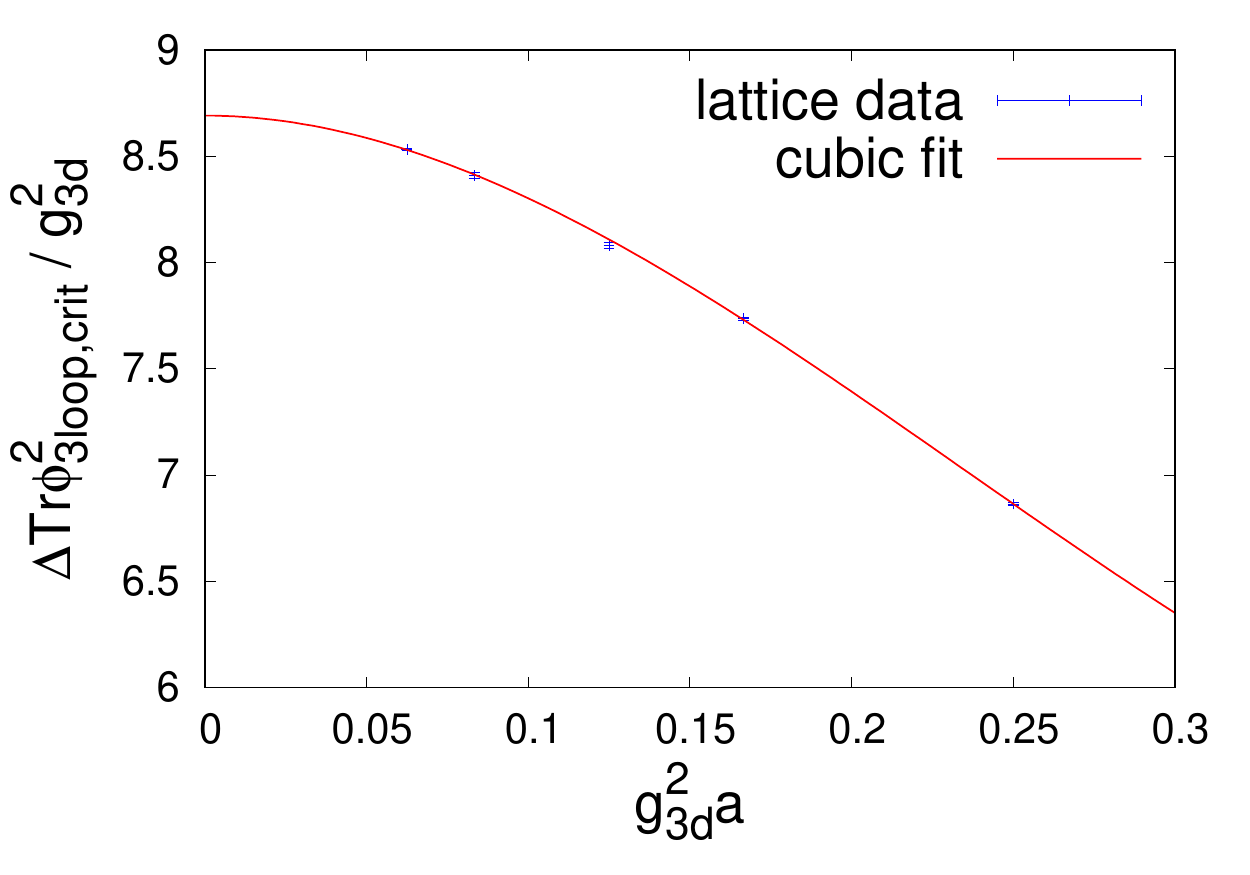} \\
		(a) $x=0.0463597$ & (b) $x=0.0677528$ \\
		\includegraphics[scale=0.5]{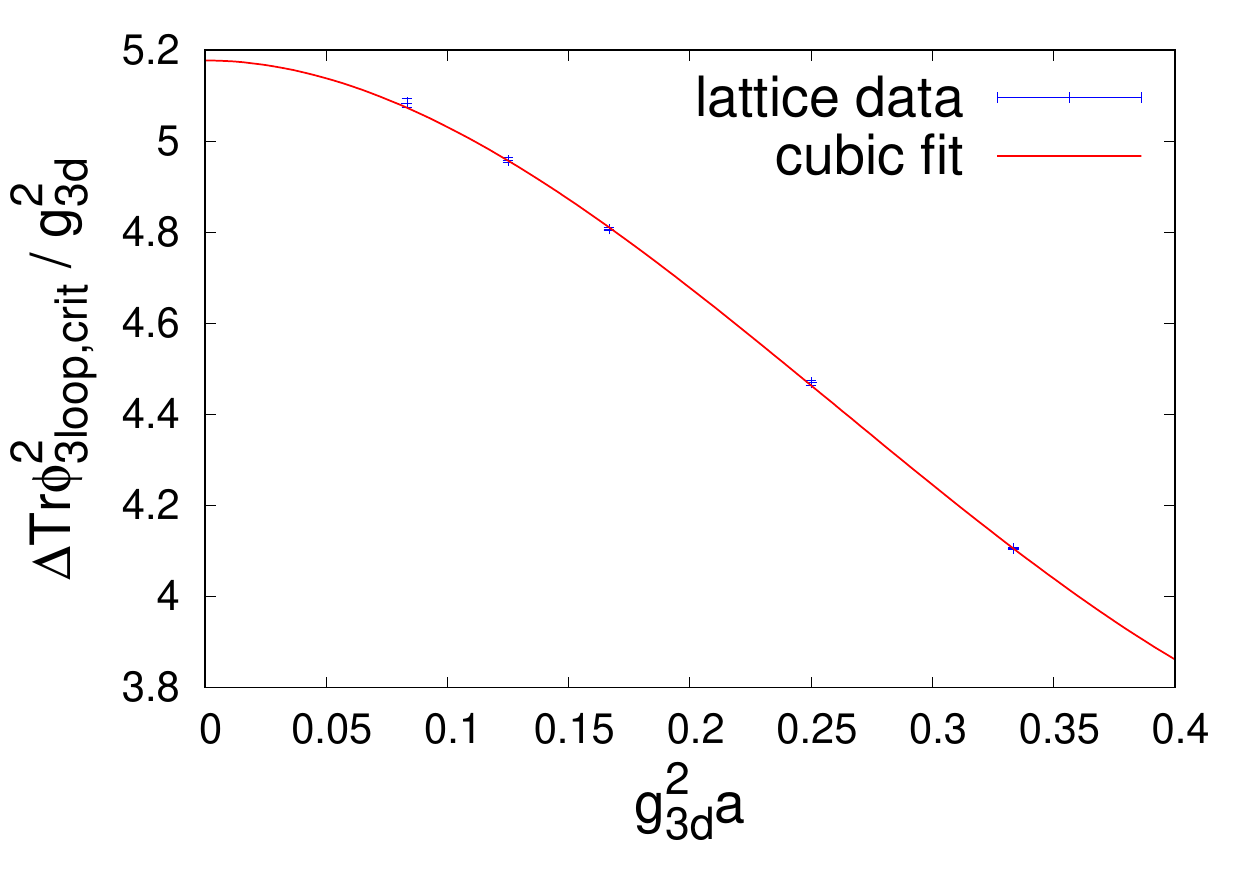} &
		\includegraphics[scale=0.5]{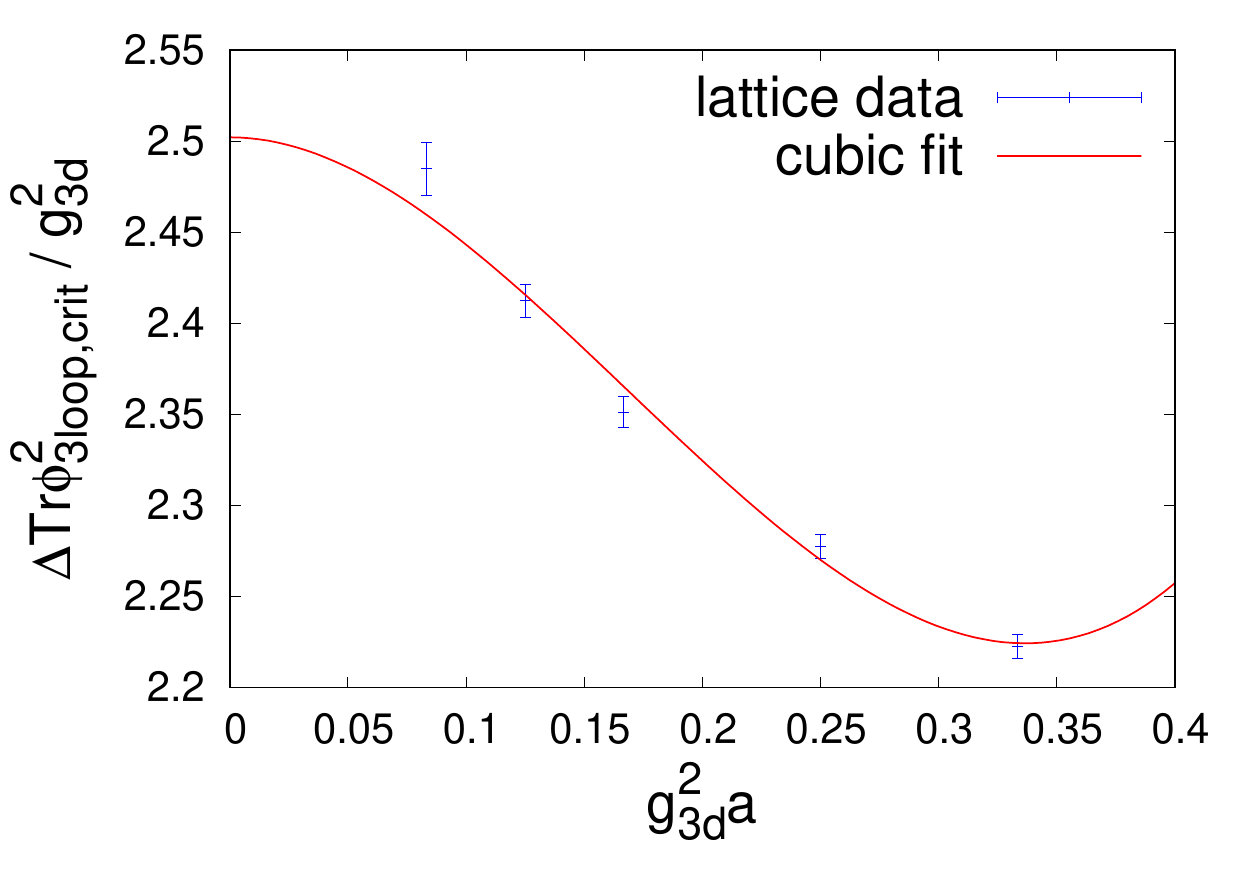} \\ 
		(c) $x=0.08896$ & (d) $x=0.13$ \\
		\includegraphics[scale=0.5]{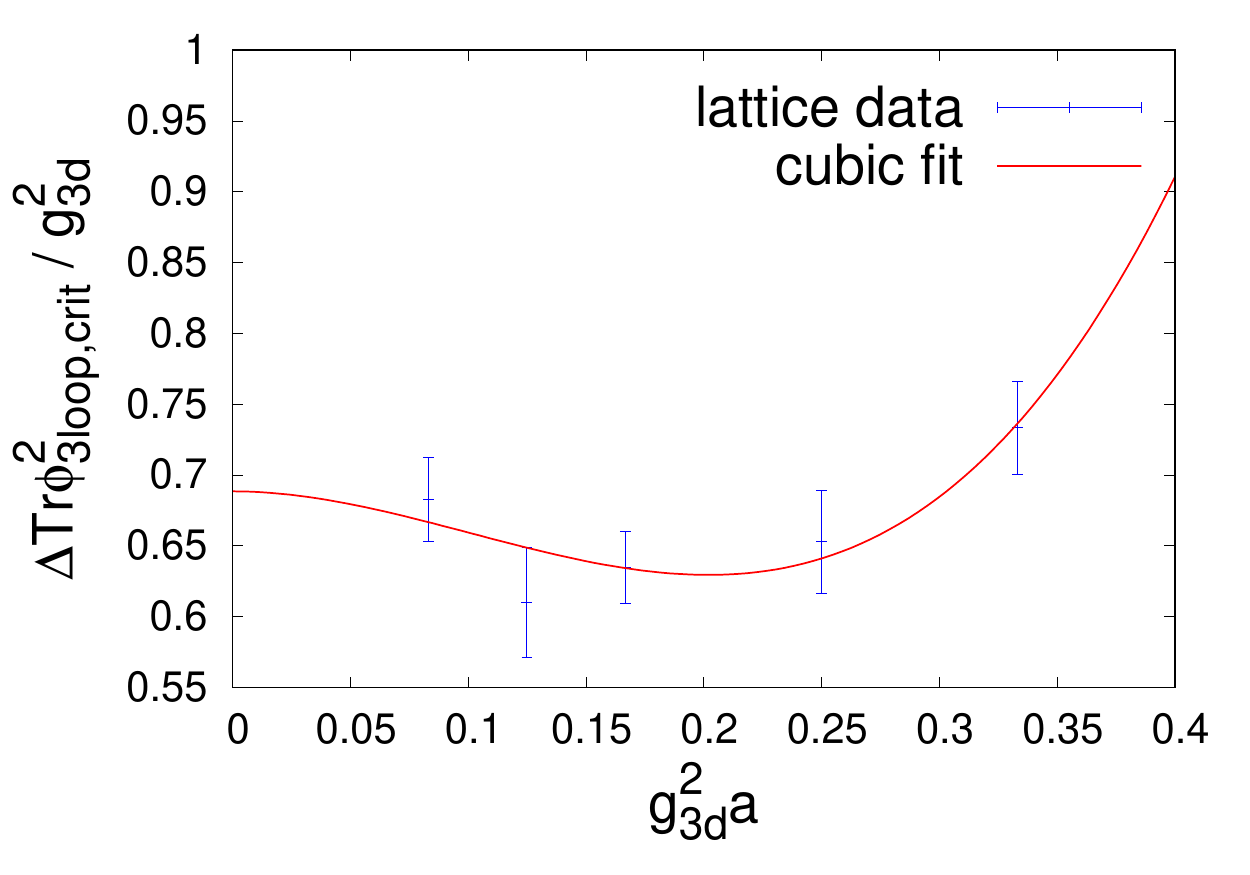} &
		\includegraphics[scale=0.5]{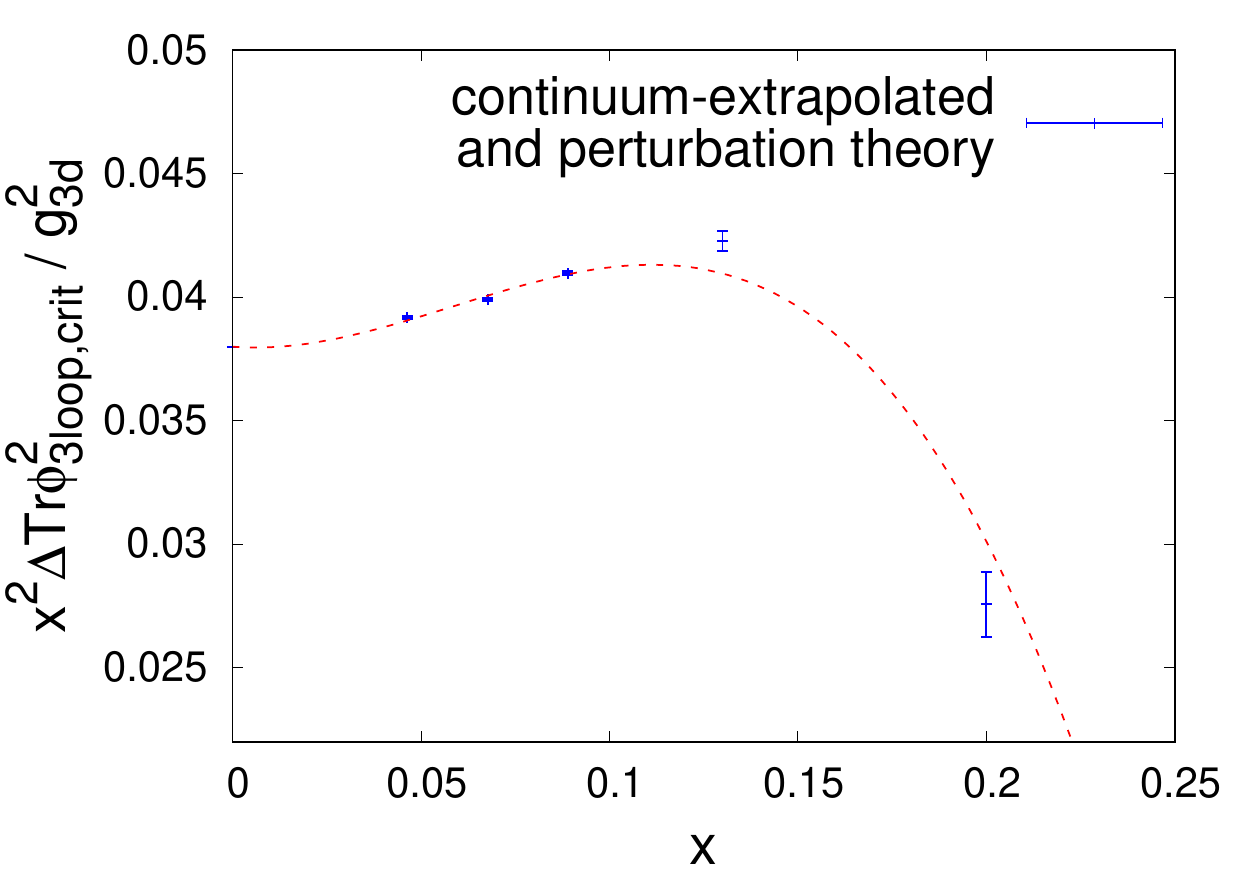}
                \\ 
		(e) $x=0.2$ & (f) $\frac{\Delta \Tr \, \Phi ^2}{\gsq}(x)$ \\

	\end{tabular}
	\caption{Continuum extrapolation of $\delphisqcont$, 
	the difference of the broken and symmetric phase value of $\phisqcont$, at 
	different $x$. The intercept in (f) was determined analytically in 
	\cite{Kajantie:1998yc} and was incorporated into that plot.}
	\label{fits_del_phisq}
\end{figure}

\begin{figure}[htbp!] 
\centering 
	\begin{tabular}{cc}
		\includegraphics[scale=0.5]{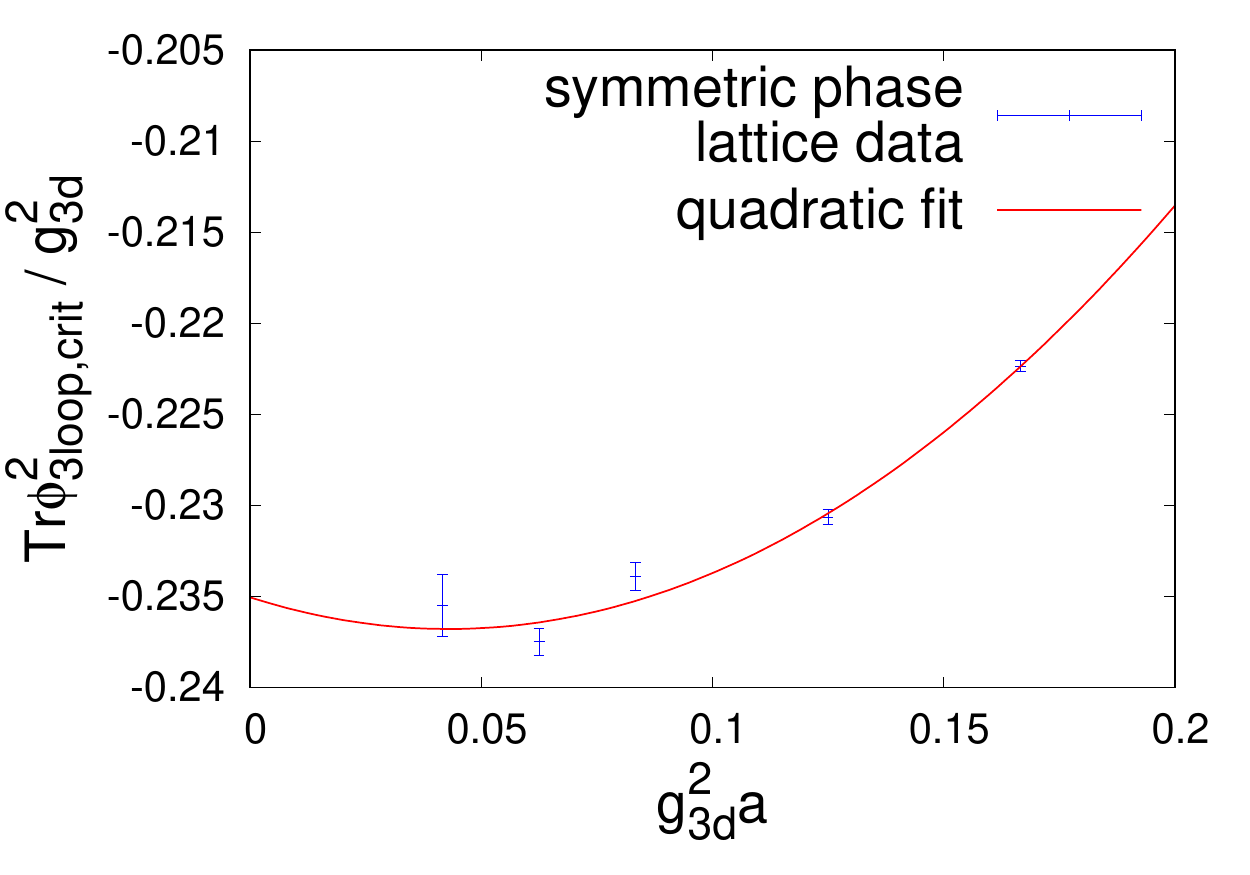} & 
		\includegraphics[scale=0.5]{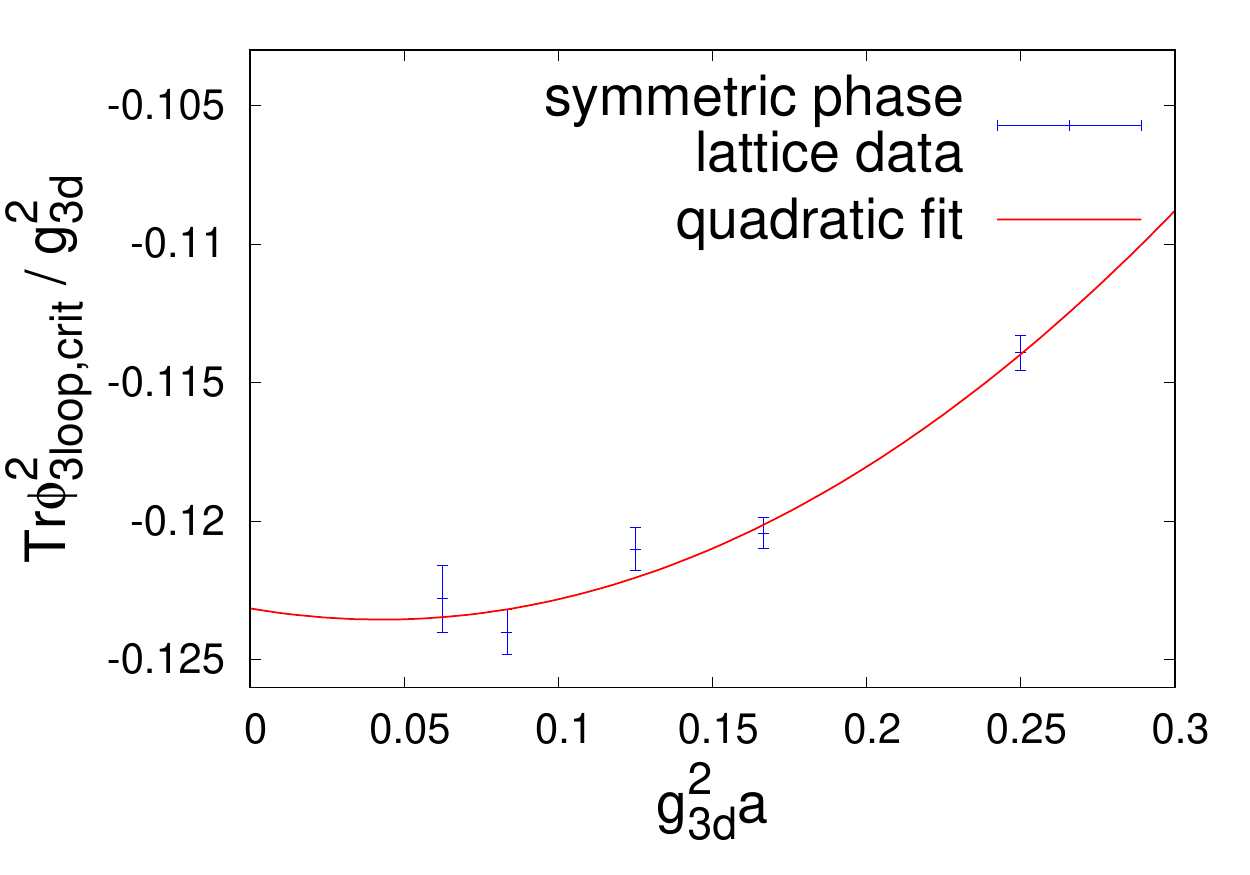} \\
		(a) $x=0.0463597$ & (b) $x=0.0677528$ \\
		\includegraphics[scale=0.5]{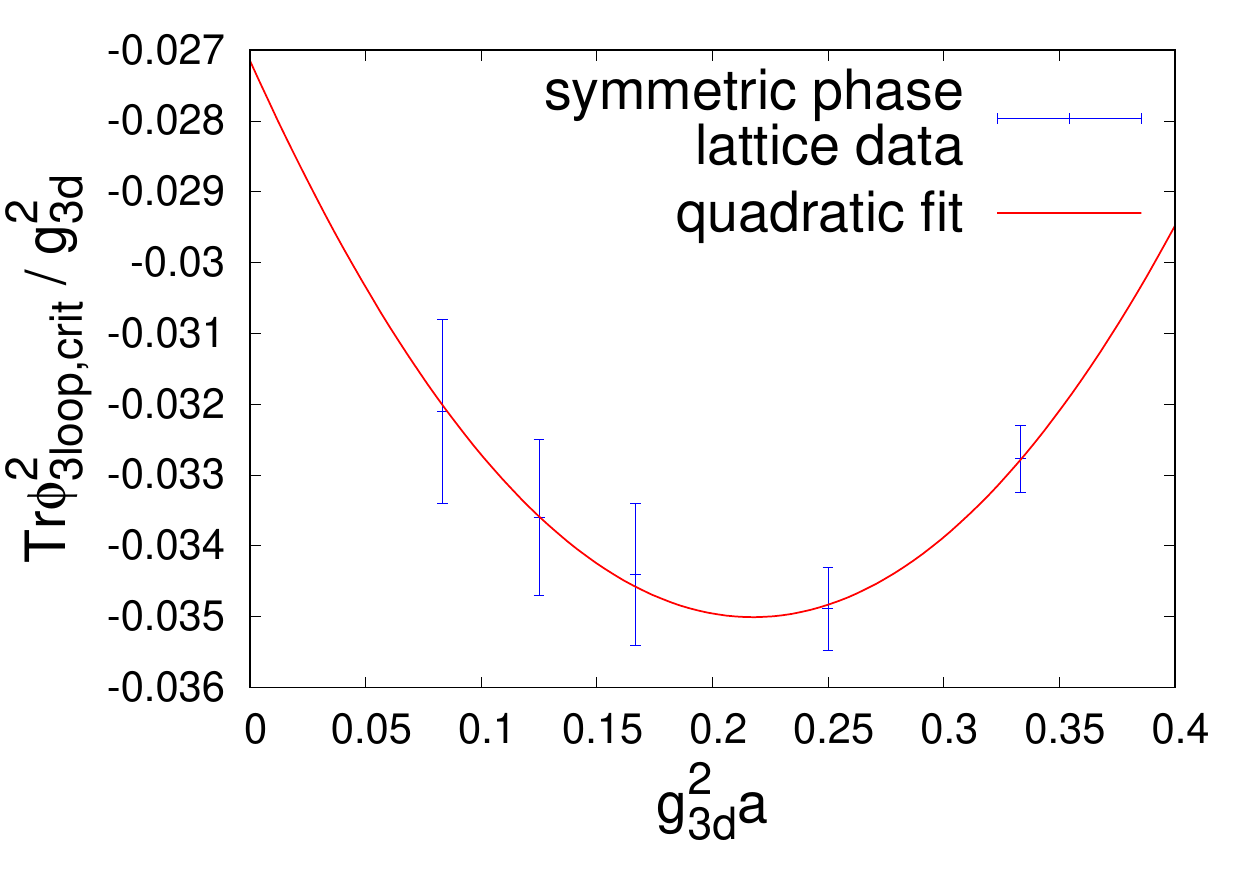} &
		\includegraphics[scale=0.5]{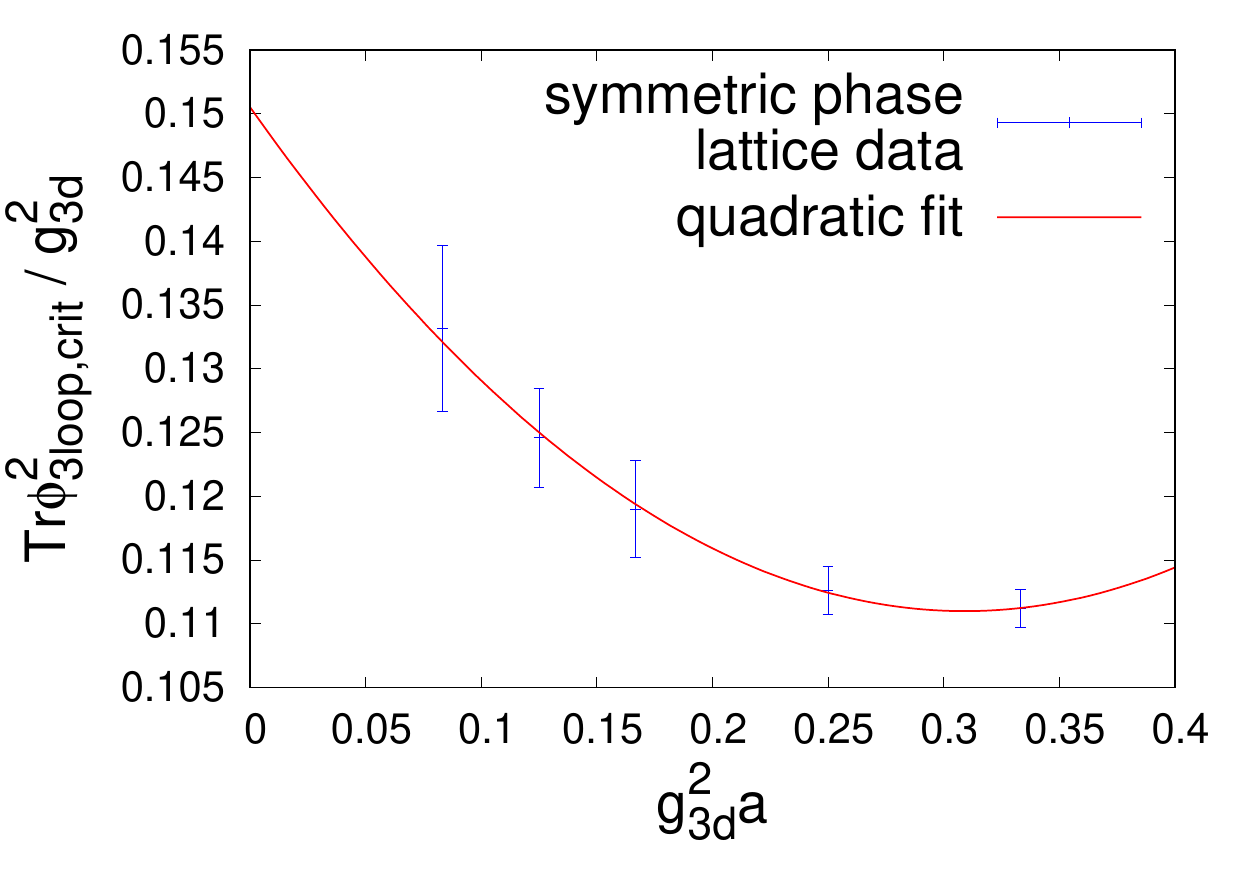} \\ 
		(c) $x=0.08896$ & (d) $x=0.13$ \\
		\includegraphics[scale=0.5]{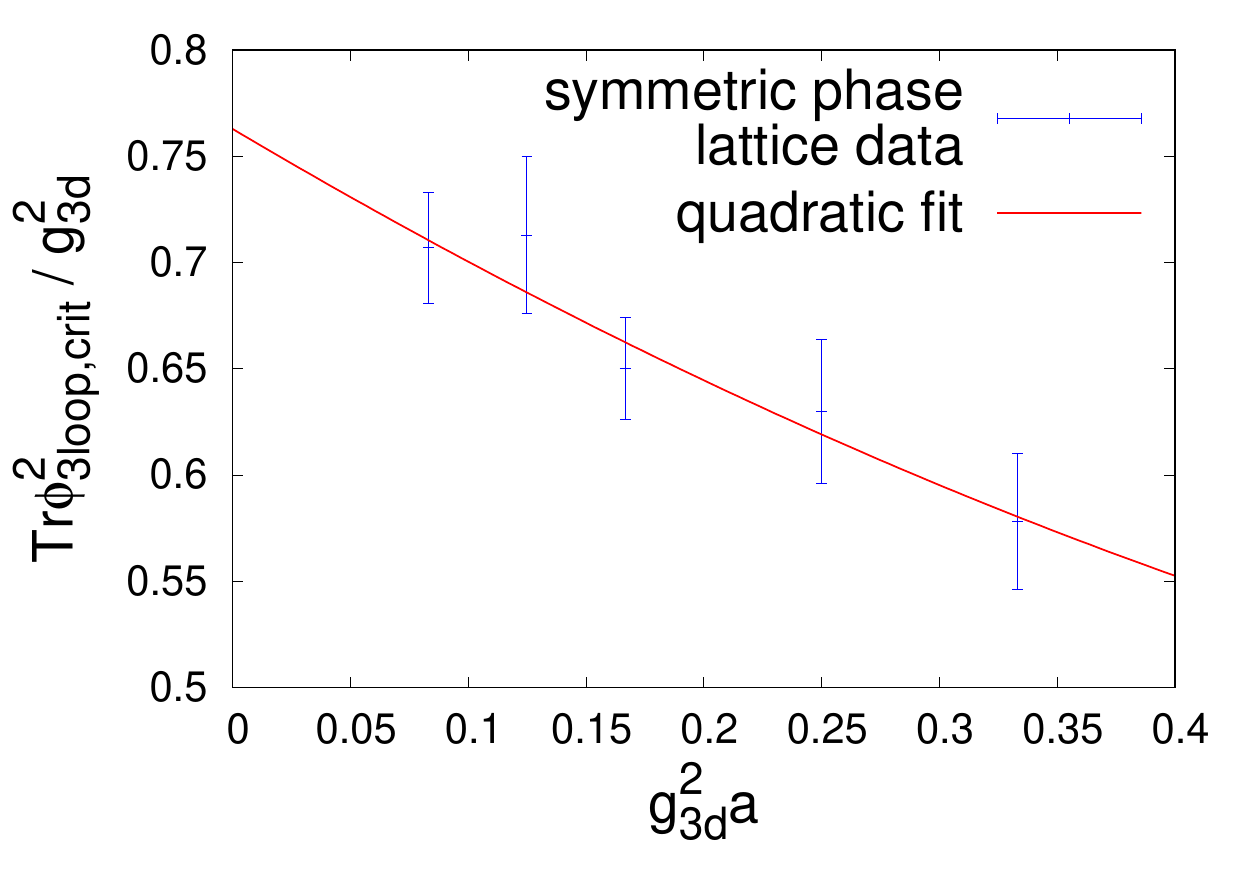} &
                \includegraphics[scale=0.5]{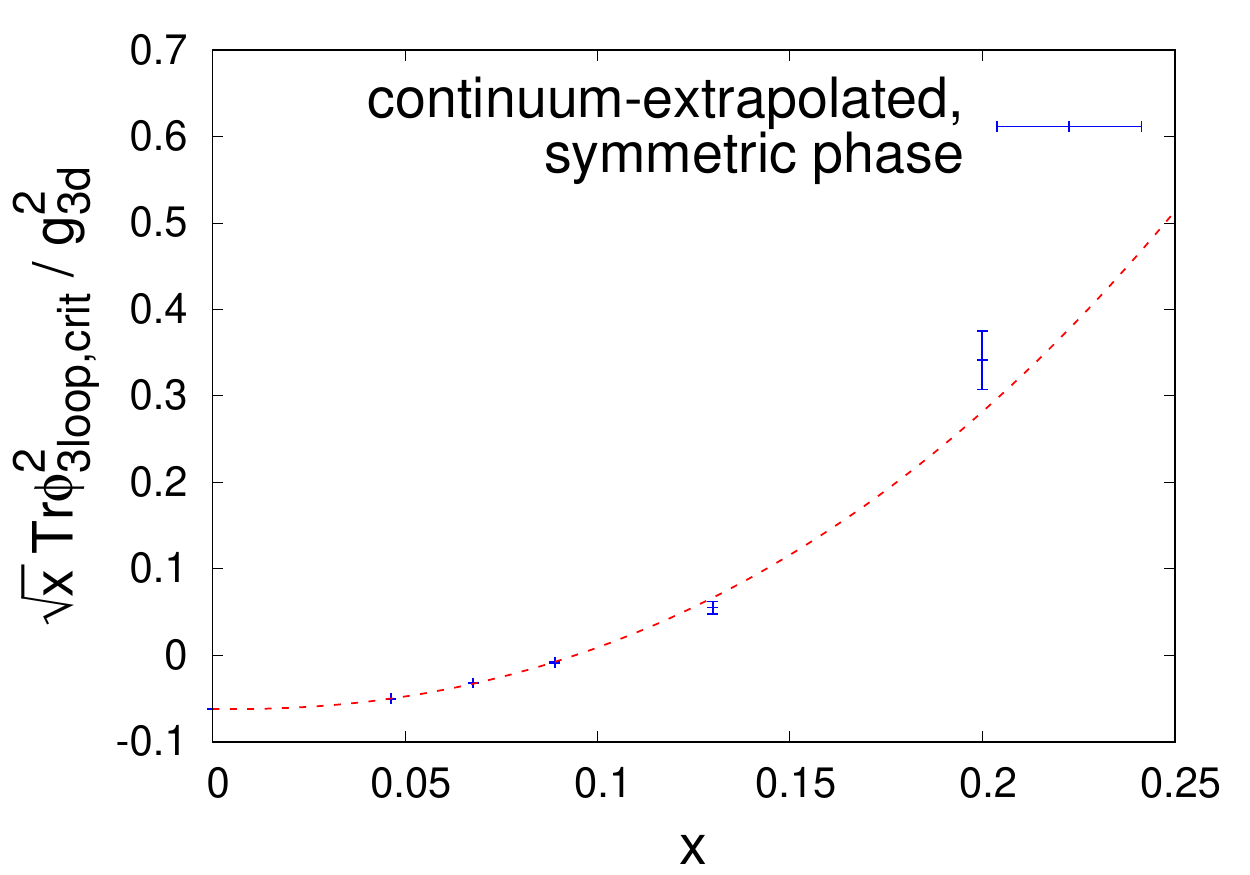} \\	
		(e) $x=0.2$ & (f) Function-of-$x$ \\
	        \includegraphics[scale=0.5]{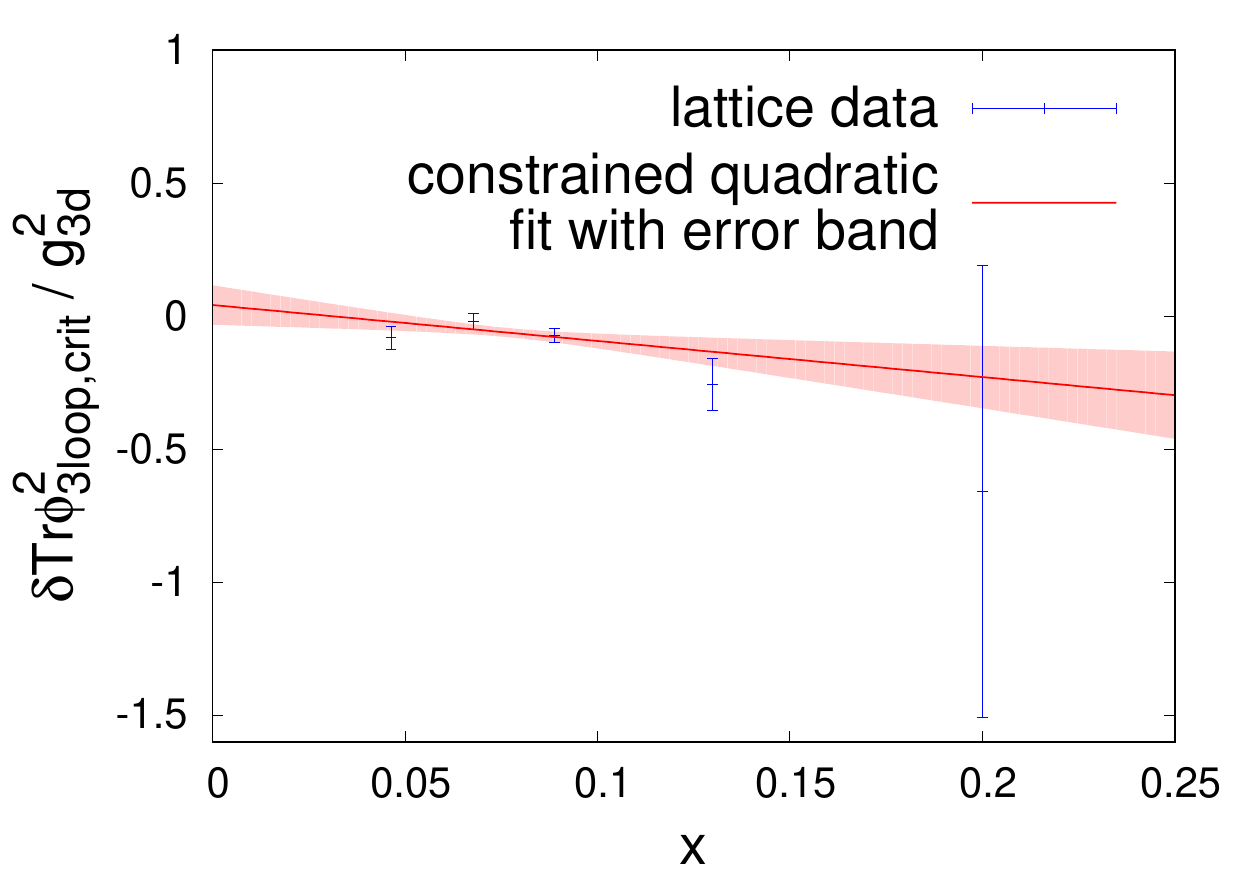}
                & \\
                (g) Grand fit $\frac{\delta \Tr \,
                  \Phi^2_\mathrm{3loop}} {\gfour a}(x)$ \\
	\end{tabular}
	\caption{Fits of $\phisqcont(\gsqa)$ at different $x$ and grand fit.}
	\label{fits_phisq_symm}
\end{figure}

With the critical values of $\phisqcont$ in both phases from Table
\ref{EQCD_sim_params} in hand, we can infer further interesting features of the 
first order phase transition. Having several values of $\phisqcont$ at the same 
physical $x$ and different lattice spacings $\gsqa$, we extrapolate both
the difference between phases $\delphisqcont$ and the symmetric phase value
$\phisqcontsymm$
to the continuum.  We provide the former in Fig.~\ref{fits_del_phisq}
and the latter in Fig.~\ref{fits_phisq_symm}.  The continuum limits,
and the linear coefficient in the fit for the case of
$\phisqcontsymm$, are provided in Table \ref{phisq_sim_results}.

\begin{table}[htbp!] 
\centering
\begin{tabular}{|c|c|c|c|}	
\hline $x$ & $\Tr \, \Phi^2_\mathrm{symm,cont}/\gsq$  & $\delta \Tr \, \Phi^2_
\mathrm{3loop} / \gfour  a$ & $\Delta \Tr \, \Phi^2_\mathrm{cont} / 
\gsq$  \\
\hline
$0.0463597$ & $-0.2350(24)$ & $-0.081(44)$ & $18.232(27)$ \\ 
$0.0677528$ & $-0.1231(22)$ & $-0.019(30)$ & $8.692(13)$ \\ 
$0.08896$ & $-0.0272(27)$ & $-0.072(25)$ & $5.177(11)$ \\ 
$0.13$ & $0.151(11)$ & $-0.256(70)$ & $2.502(24)$ \\ 
$0.2$ & $0.763(76)$ & $-0.66(85)$ & $0.689(33)$ \\
\hline
\end{tabular}
\caption{Continuum-extrapolated $\phisqcont$ in the symmetric phase at 
criticality, $\OO(a)$ operator improvement of $\phisqcont$ and continuum-
extrapolated difference of $\phisqcont$ in the two phases at criticality.}
\label{phisq_sim_results}
\end{table}

The limiting values of the fits provide us with two interesting pieces
of information about the phase transition in this theory.  The most
interesting is $\delphisqcont$, which measures the strength of the
phase transition.  For small $x$ we can predict this strength
perturbatively; the limiting behavior is \cite{Kajantie:1998yc}
$\delphisqcont = 3/(8\pi^2 x^2)$,
which we also include in the last frame of Fig.~\ref{fits_del_phisq}.
We have provided a cubic fit to guide the eye, but it should not be
taken seriously; the strength of the phase transition is a
nonperturbative quantity and there is no reason to expect it to take
such a simple form.  In fact, we know that $\delphisqcont \to 0$ as
$x \to x_{\mathrm{triple}}$, with a nontrivial critical exponent.
Note that in the fit for the $a$ dependence of $\delphisqcont$, we
have fitted to a polynomial without a linear term; this is because the
known $\OO(a)$ corrections are sufficient to eliminate such a linear
correction in the \textsl{difference} between phases of $\phisqcont$.

\subsection{Additive operator improvement}

On the other hand, the value of either $\phisqcontsymm$ or
$\phisqcontbrok$, by themselves, still contain $\OO(a)$ errors, since
there is an unknown additive renormalization to the operator
$\Tr\,\Phi^2$ which arises at 3 loops.  Since the correction is
additive and both phases were explored at the same $y$ value, these
effects cancel in the difference.  We took advantage of this
cancellation in the last subsection.  But now our goal is to use this
linear behavior to extract the unknown $\OO(a)$ additive corrections
to the $\Phi^2$ operator.  These arise at 3 loops in a perturbative
lattice-continuum matching calculation, which is prohibitive; so we
will again try to extract them from the data.

\begin{figure}
  \includegraphics[scale=0.41]{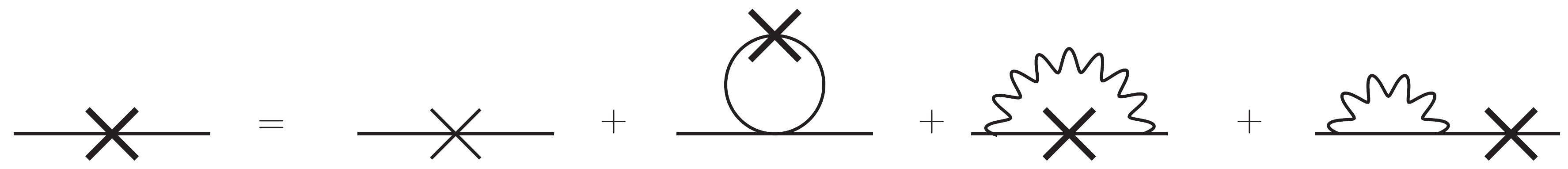}
  \caption{Diagrams generating $Z_m$ the
    multiplicative $\Tr\:\Phi^2$ operator renormalization.  Heavy
    crosses are the renormalized operator (with $Z_m$ factor), while
    light crosses are the bare operator.}
    \label{fig:resum}
\end{figure}

Because we are working to 3 loops, we must specify quite carefully how
1-loop multiplicative effects will be implemented, since they can
multiply one and two loop additive effects to give 3-loop level
contributions, which then differ depending on our exact procedure.
Here we will depart slightly from the procedure of Refs
\cite{Moore:1997np,Arnold:2001ir,DOnofrio:2014mld}.  We write the
continuum expectation value as
\begin{equation}
  \label{phi_contlatt}
  \Tphicont = Z_m \left( Z_\Phi \Tphilat - \frac{\delta \Phi^2}{\gsq} \right) \,,
\end{equation}
where $Z_m$ is the 1-loop multiplicative renormalization
factor of the $\Tr\,\Phi^2$ operator, and $Z_\Phi$ accounts for our
choice of scalar field normalization on the lattice (see
Eq.~(\ref{latthopping})).  Examining the
3-loop diagrams, we find that certain 3-loop effects are absorbed if
we define $Z_m$, \textsl{resumming} the Dyson
series.  That is, in Figure~\ref{fig:resum}, we take the operator
inserted on the 1-loop diagrams to be the resummed, rather than the
bare, operator, which will sum the Dyson series, leading to an
expression for $Z_m^{-1}$.
Slightly rearranging Eq.(32,33) of Ref.~\cite{Moore:1997np}, we find
\begin{equation}
  Z_m^{-1} = 1 + \frac{\gsqa}{4\pi} \left(
  3N \xi + \frac{N \Sigma}{6} - (N^2+1) \xi x \right) \,.
\end{equation}
Here $\xi = 0.152859324966$ and $\Sigma = 3.17591153562522$ are
standard integrals encountered in the 1-loop lattice-continuum
matching.  We are also writing the number of colors $N=3$ explicitly,
to show the detailed dependence on the number of colors.
With this definition, the two-loop and partially 3-loop
result for $\delta \Phi^2$ is
\begin{eqnarray}
  \label{phi3loop}
  \frac{\delta \Phi^2}{\gsq} & = & \frac{N^2-1}{2 \gsqa} \left[
    \frac{\Sigma}{4\pi} - \frac{\xi y g_{3\mathrm{d}}^4 a^2}{4\pi}
    + \frac{N \gsqa}{(4\pi)^2} Z_m^{-1}
    \left( 2 \ln \frac{6}{\gsqa} + 2\zeta - 2\delta
    + \frac{\Sigma^2}{2} \right) \right.
    \nonumber \\ && \phantom{\frac{N^2-1}{2 \gsqa}} \left. {}
    + \frac{g_{3\mathrm{d}}^4 a^2}{(4\pi)^3}
    \Big( 2\xi (N^2+1)(x^2 - N x) \ln(\gsqa) 
    + C_{\Phi a} + C_{\Phi b} x + C_{\Phi c} x^2    \Big) \right] . \quad
\end{eqnarray}
The unknown coefficients $C_{\Phi a}$, $C_{\Phi b}$ and
$C_{\Phi c}$ capture the remaining $\OO(a)$ corrections from 3-loop
diagrams which are not iterations of simpler 1 and 2 loop diagrams.
Note that Eq.~(\ref{phi3loop}) contains several terms proportional to
$\ln(\gsqa)$.  These arise from logarithmic divergences in the
continuum theory, regulated at our choice of renormalization point
$\mu=\gsq$ but then cut off on the lattice at the scale $1/a$.  The
term next to $\zeta-\delta$ is the explicit $\mu^2$ dependence of
$\Phi^2/\gsq$ and is therefore expected; the coefficient $Z_m^{-1}$
ensures that it enters Eq.~(\ref{phi_contlatt}) with precisely the
right continuum normalization.  The log terms proportional to $\xi$ in
the last line cancel the $\mu$ dependence of the mass squared ($y$) in
the mass-dependent $\OO(a)$ shift in the first line.

It remains to determine the three coefficients in the last line.
In fact, we only need to fit two of these coefficients; one of them,
$C_{\Phi c}$, represents pure-scalar corrections, which
can be extracted from Ref.~\cite{Arnold:2001ir}.  The reference
performs the calculation for an improved hopping term, but repeating the
calculation for the nearest-neighbor hopping term we use here%
\footnote{The rest of the $\OO(a)$ corrections are not known for
  improved actions, which is why we do not attempt to use an improved
  action here.  Using improved actions only really helps if one can
  complete the 2-loop $\OO(a^2)$ matching calculation; this is
  feasible in a scalar theory, but does not appear practical in a
  gauge theory as we consider here.},
we find that
\begin{equation}
  C_{\Phi c} =  2(N^2+1) C_4 \,,
  \qquad
  C_4 = 0.5630(4) \,,
\end{equation}
which differs from the result in the reference,
$C_{4,\mathrm{Ref}}=0.2817$, because of the different scalar
dispersion between the nearest-neighbor hopping term used here and the
improved hopping term used there.%
\footnote{%
  Specifically, in passing from the improved to the unimproved hopping
  term, Ref.~\cite{Arnold:2001ir} (B38) has $0.30837 \to 0.2268854$;
  (B40) has $.00031757 \to .000490546$, the value of $\xi$ changes
  from $\xi = -.08365\to +.1528593$, and (B41) changes from
  $.0985(6) \to .1063(4)$.}

After performing these subtractions, we can fit the residual
linear $a$-dependence of $\Tphilat$ for each $x$ value we consider, and
extract the coefficients $C_{\Phi a},C_{\Phi b}$ from a grand fit in
complete analogy with the $m^2$ effects we consider in the main text.
We find 
\begin{align}
\label{Cphi2}
C_{\Phi a} &= (-21 \pm 37) \\ 
C_{\Phi b} &= (6.7 \pm 4.7) \times 10^{2} \, ,
\end{align}
again by constrained curve fitting. The grand fit is displayed
in the seventh frame of Fig.~\ref{fits_phisq_symm}.
Unfortunately, it appears that our results fail to constrain these
coefficients very much. The full covariance matrix can be found in Tab.~\ref{cov_mat_phisq_fit}. From the covariance matrix, we can also see that the error of $C_{\Phi c}$ is by far the smallest, so value and error of $C_{\Phi c}$ were not changed by the constrained fit.

\begin{table}
\centering
\begin{tabular}{|c|c c c| }	
\hline
$\mathrm{cov}(C_i,C_j)$ & $C_{\Phi a}$ & $C_{\Phi b}$ & $C_{\Phi c}$   \\
\hline
$C_{\Phi a}$ & $21694$ & $-267617$ & $5.3859 \cdot 10^{-6}$ \\
$C_{\Phi b}$ & $-267617$ & $3.4914 \cdot 10^{6}$  & $-1.432 \cdot 10^{-4}$ \\
$C_{\Phi c}$ & $5.3859 \cdot 10^{-6}$ & $-1.432 \cdot 10^{-4}$ & $9.0 \cdot 10^{-4}$ \\
\hline
\end{tabular}
\caption{Covariance matrix of the grand fit of the missing 3 loop $\frac{\delta \Phi^2}{\gsq}$-contribution.}
\label{cov_mat_phisq_fit}
\end{table}

\bibliographystyle{unsrt}
\bibliography{references}

\end{document}